%% file: B2pi.tex
\documentstyle[preprint,prd,aps,eqsecnum,psfig]{revtex}
\tightenlines
\draft

\newcommand{\boldvec}[1]{\mbox{\boldmath$#1$}}
\newcommand{\backvec}[1]{\stackrel{\leftarrow}{#1}}

\begin{document}

\date{\today}
\preprint{
  $
  \begin{array}{r}
    \mbox{HUPD-0105}\\
    \mbox{YITP-01-48}\\
    \mbox{KEK-CP-109}
  \end{array}
  $
  }
\title{Differential decay rate of 
       $B \rightarrow \pi l \nu$ semileptonic decay 
       with lattice NRQCD}
\author{
  S.~Aoki$^{\rm a}$,
  M.~Fukugita$^{\rm b}$,
  S.~Hashimoto$^{\rm c}$,
  K.-I.~Ishikawa$^{\rm c}$
  N.~Ishizuka$^{\rm a,d}$,
  Y.~Iwasaki$^{\rm a,d}$,
  K.~Kanaya$^{\rm a,d}$,
  T.~Kaneko$^{\rm c}$,
  Y.~Kuramashi$^{\rm c}$,
  M.~Okawa$^{\rm c}$,
  T.~Onogi$^{\rm e,f}$,
  S.~Tominaga$^{\rm d}$,
  N.~Tsutsui$^{\rm c}$,
  A.~Ukawa$^{\rm a,d}$,
  N.~Yamada$^{\rm c}$
  T.~Yoshie$^{\rm a,d}$
  }
\address{
  $^{\rm a}$Institute of Physics,
  University of Tsukuba,
  Tsukuba 305-8571, Japan\\
  $^{\rm b}$Institute for Cosimic Ray Research,
  University of Tokyo,
  Kashiwa 277-8582, Japan\\
  $^{\rm c}$High Energy Accelerator Research Organization (KEK),
  Tsukuba 305-0801, Japan\\
  $^{\rm d}$Center for Computational Physics,
  University of Tsukuba,
  Tsukuba 305-8571, Japan\\
  $^{\rm e}$Department of Physics, 
  Hiroshima University,
  Higashi-Hiroshima 739-8526, Japan\footnote{address before April, 2001}\\
  $^{\rm f}$ Yukawa Institute for Theoretical Physics,
  Kyoto University, Kyoto 606-8502, Japan
  \footnote{address since April, 2001}
  }
\maketitle

\begin{abstract}
  We present a lattice QCD calculation of 
  $B\rightarrow \pi l \nu$ semileptonic decay form factors
  in the small pion recoil momentum region.
  The calculation is performed on a quenched $16^3 \times 48$
  lattice at $\beta=5.9$ with the NRQCD action including the
  full $1/M$ terms. 
  The form factors 
  $f_1(v\cdot k_{\pi})$ and $f_2(v\cdot k_{\pi})$
  defined in the heavy quark effective theory 
  for which the heavy quark scaling is manifest are adpoted, 
  and we find that the $1/M$ correction to the scaling is small for 
  the $B$  meson. 
  The dependence of form factors on the light quark mass
  and on the recoil energy is found to be mild, and we use a
  global fit of the form factors at various quark masses and
  recoil energies to obtain model independent results for
  the physical differential decay rate.
  We find that the $B^*$ pole contribution dominates the
  form factor $f^+(q^2)$ for small pion recoil energy, and
  obtain the differential decay rate integrated over the
  kinematic region $q^2 >$ 18~GeV$^2$ to be 
  $|V_{ub}|^2 \times ( 1.18 \pm 0.37 \pm 0.08 \pm 0.31 )$
  psec$^{-1}$, where the first error is statistical,
  the second is that from perturbative calculation, and the
  third is the systematic error from finite lattice spacing
  and the chiral extrapolation. 
  We also discuss the systematic errors in the soft pion
  limit for $f^0(q^2_{max})$ in the present simulation.
\end{abstract}

\input{1.Introduction.tex}

\input{2.Form_Factors.tex}

\input{3.Lattice_NRQCD.tex}

\input{4.Matching.tex}

\input{5.Lattice_calculation.tex}

\input{6.Results.tex}

\input{7.Comparison.tex}

\input{8.Phenomenology.tex}

\input{9.Conclusion.tex}

\section*{Acknowledgment}
We thank Andreas Kronfeld for useful discussion and for
providing us with their numerical data for comparison.
This work is supported by the Supercomputer Project No.66
(FY2001) of High Energy Accelerator Research Organization
(KEK), and also in part by the Grants-in-Aid of the Ministry
of Education (Nos. 10640246, 11640294,
12014202, 12640253, 12640279, 12740133, 13640260 and 13740169).
K.-I.I. and N.Y. are supported by the JSPS Research
Fellowship.

\input{Reference.tex}

\input{Table.tex}
\input{Figure.tex}

\end{document}

%% file: 1.Introduction.tex
\section{Introduction}
\label{sec:Introduction}

The exclusive decay modes
$B^0\rightarrow\pi^-l^+\nu_l$ and
$B^0\rightarrow\rho^-l^+\nu$ 
may provide us with the best experimental input to determine  
the Cabibbo-Kobayashi-Maskawa (CKM) matrix element
$|V_{ub}|$.
At present these decays are measured by CLEO
\cite{CLEO_96,CLEO_00} with error of order 20\%.
A prerequisite for the determination of $|V_{ub}|$ is an
accurate calculation of the form factors involved in these
semileptonic decays, but the theoretical prediction of the
form factors for the entire kinematical range is still
difficult. 
However, with the advent of the B factories, BaBar, Belle
and CLEO~III, we expect that the differential decay rate
will be measured precisely as a function of momentum
transfer $q^2$ in near future.
This means that to determine $|V_{ub}|$ we do not
necessarily need the form factor for the entire kinematic
region of $q^2$, but calculations in a certain limited range
of $q^2$ will practically suffice.

Lattice QCD provides a promising framework to compute the
form factors without resorting to specific phenomenological
models.
Exploratory studies have already been made by a few groups 
\cite{Bernard_lat00,Hashimoto_lat99,Flynn_Sachrajda_HF2},
but more extensive studies are clearly needed to provide
realistic predictions. In this work we attempt to compute
the form factors and differential decay rates of
$B\rightarrow\pi l\nu$ for the momentum range 
$q^2>$ 18~GeV$^2$, which is set by the condition that the
spatial momenta of the initial and final hadrons be much
smaller than the lattice cutoff $1/a$,
$|\mathbf{k}| \ll$ $1/a \simeq$ 2~GeV/c,
to avoid discretization error. 

An important point in the calculation of the $B$ meson
matrix elements is to reduce the systematic error arising
from a heavy quark mass $M$ which is larger than $1/a$.
One approach adopted in the literature is to calculate the
matrix elements with a relativistic action for heavy quarks
around the charm quark mass and to extrapolate them to the
bottom quark mass.
Although this approach seems to work reasonably well in the recent
studies of $ B\rightarrow\pi l\nu$ form factors
\cite{UKQCD_00,APE_00}, the systematic error is 
magnified in the extrapolation and the heavy quark mass
dependence would not be correctly predicted. 
This problem can be avoided by using a variant of the Heavy
Quark Effective Theory (HQET), in which the 
the heavy quark is treated non-relativistically.

A natural implementation of the idea of HQET on the lattice
is the non-relativistic QCD (NRQCD) \cite{NRQCD}, which we
employ in this work.
With the NRQCD action the heavy quark mass dependence of the
form factors can be reliably calculated
\cite{Hashimoto_et_al_98}, since the action is written as an
expansion in terms of inverse heavy quark mass and higher
order terms can optionally be included to achieve desired
accuracy.
In the $B\rightarrow\pi l\nu$ decay near zero recoil of the pion, we
find that the heavy quark expansion converges well at the
next-to-leading order in $1/M$. 

An alternative implementation of the HQET is the Fermilab
formalism \cite{El-Khadra_Kronfeld_Mackenzie_97}, in which
results from the conventional relativistic lattice action 
are reinterpreted in terms of a non-relativistic effective Hamiltonian.
This formalism shares an advantage similar to that of NRQCD, 
and has recently been applied to a $B\rightarrow\pi l\nu$ decay 
calculation \cite{Fermilab_01}.

In the application of the HQET to the $B\rightarrow\pi l\nu$
decay, it is more natural to work with the form factors 
$f_1(v\cdot k_{\pi})$ and $f_2(v\cdot k_{\pi})$
\cite{Burdman_Ligeti_Neubert_Nir_94}, 
where $v^{\mu}$ is a heavy quark velocity and $k_{\pi}^{\mu}$ 
is a four-momentum of pion, rather than the conventional 
$f^+(q^2)$ and $f^0(q^2)$.
This is because the argument $v\cdot k_{\pi}$, which is the
energy of the pion in $B$ meson rest frame, is well-defined
in the limit of infinitely heavy quark mass, and the heavy
quark scaling, 
\textit{ i.e.}
$f_{1,2}(v\cdot k_{\pi})\rightarrow$ constant
as $M \rightarrow \infty$,
 is manifest in the new set of the form factors.

We calculate $f_{1,2}(v\cdot k_{\pi})$ using the NRQCD action
on a quenched lattice of size $16^3\times 48$ at $\beta=5.9$
corresponding to $1/a\approx 1.6$ GeV.
The action we use includes the full
terms of order $1/M$. 
The $O(a)$-improved Wilson fermion action is used for the light
quark. We prepare a large statistical sample, accumulating 
2,150 gauge configurations to reduce 
statistical noise which becomes large for states
with finite momenta. This enables us to obtain
good signals for the form factors for a finite spatial
momentum of the pion.

This paper is organized as follows.
In the next section we briefly review the definition of the
HQET motivated form factors $f_{1,2}(v\cdot k_{\pi})$ of
Burdman \textit{et al.} \cite{Burdman_Ligeti_Neubert_Nir_94}
and their relation to the conventional form factors.
We summarize the definition of the NRQCD action 
in Section~\ref{sec:Lattice_NRQCD}, and discuss
matching of the
heavy-light vector current on the lattice with that in the continuum in
Section~\ref{sec:Matching_of_the_heavy-light_current}. 
We describe our lattice calculation in
Section~\ref{sec:Lattice_calculation}, and the results are
presented in
Section~\ref{sec:Results_for_the_form_factors}. 
Section~\ref{sec:Comparison_with_other_calculations} is
given to a comparison with other lattice calculations, and 
phenomenological implications are discussed in
Section~\ref{sec:Phenomenological_Implications}.
Our conclusions are presented in
Section~\ref{sec:Conclusions}. 


%% file: 2.Form_Factors.tex
\section{The HQET form factors for $B\rightarrow\pi l \nu$}
\label{sec:HQET_form_factors}

The matrix element 
$\langle\pi(k_{\pi})|\bar{q}\gamma_{\mu}b|B(p_B)\rangle$ 
for the heavy-to-light semileptonic decay $B\rightarrow\pi l\nu$ is
usually parameterized as
\begin{equation}
  \label{eq:f+f0}
  \langle\pi(k_{\pi})|\bar{q}\gamma^{\mu}b|B(p_B)\rangle
  = f^+(q^2) 
  \left[ 
    (p_B+k_{\pi})^{\mu} 
    - \frac{m_B^2-m_{\pi}^2}{q^2} q^{\mu}
  \right]
  + f^0(q^2) \frac{m_B^2-m_{\pi}^2}{q^2} q^{\mu},
\end{equation}
with $p_B$ and $k_{\pi}$ the momenta of the initial and final
pseudo-scalar mesons and $q = p_B-k_{\pi}$.
When lepton mass is negligible, the
momentum transfer $q^2$ ranges from 0 to
$q^2_{max}=(m_B-m_{\pi})^2$.
From the kinematics
\begin{equation}
  \label{eq:v.k_pi}
  E_{\pi} = v\cdot k_{\pi} =
  \frac{m_B^2-m_{\pi}^2-q^2}{2m_B},
\end{equation}
where $v=p_B/m_B$ is the four-velocity of the initial $B$ meson,
a low $q^2$ corresponds to a large
recoil momentum of pion, for which the lattice calculation is
not easy.
In the other limit $q^2\sim q^2_{max}$, however, the energy of the  
pion $E_{\pi}$ in the $B$ meson rest frame is minimum,
so that spatial momenta of the initial and final hadrons are small
compared to the lattice cutoff, and the lattice calculation 
will give a reliable answer. 

In HQET, it is more natural to use $v^{\mu}$ and
$k_{\pi}^{\mu}$ as independent four-vectors rather than 
$p_B^{\mu}$ and $k_{\pi}^{\mu}$.
Burdman \textit{et al.} \cite{Burdman_Ligeti_Neubert_Nir_94}
defined the form factors 
$f_1(v\cdot k_{\pi})$ and $f_2(v\cdot k_{\pi})$ by
\begin{equation}
  \label{eq:f1f2}
  \langle\pi(k_{\pi})|\bar{q}\gamma^{\mu}b|B(v)\rangle
  = 2 \left[
    f_1(v\cdot k_{\pi}) v^{\mu} +
    f_2(v\cdot k_{\pi}) \frac{k_{\pi}^{\mu}}{v\cdot k_{\pi}}
  \right],
\end{equation}
where the heavy meson field is normalized with the factor $2v^0$
instead of the usual $2p_B^0$, so that 
$\sqrt{m_B}|B(v)\rangle = |B(p_B)\rangle$.
The new form factors are functions of $v\cdot k_{\pi}$ and
defined over the range $[m_{\pi},(m_B^2-m_{\pi}^2)/2m_B]$.
As seen from definition (\ref{eq:f1f2}) there is no
explicit dependence on the heavy meson mass.
Therefore, heavy quark scaling as $M\rightarrow\infty$
is manifest, namely, $f_{1,2}(v\cdot k_{\pi})$ become
independent of $M$ up to logarithms arising from the
renormalization of the heavy-light current.
Finite $M$ corrections are given as a
power series in $1/M$. 

The relation between the two definitions of form factors is
given by 
\begin{eqnarray}
  \label{eq:f+_from_f1f2}
  f^+(q^2) & = &
  \sqrt{m_B} \left\{
      \frac{f_2(v\cdot k_{\pi})}{v\cdot k_{\pi}} +
      \frac{f_1(v\cdot k_{\pi})}{m_B}
    \right\},
  \\
  \label{eq:f0_from_f1f2}
  f^0(q^2) & = &
  \frac{2}{\sqrt{m_B}}
  \frac{m_B^2}{m_B^2-m_{\pi}^2}
  \Biggl\{
    \left[ 
      f_1(v\cdot k_{\pi}) + f_2(v\cdot k_{\pi})
    \right]
  \nonumber\\
  & & 
  \left.
    -
    \frac{v\cdot k_{\pi}}{m_B}
    \left[ 
      f_1(v\cdot k_{\pi}) + 
      \frac{m_{\pi}^2}{(v\cdot k_{\pi})^2}
      f_2(v\cdot k_{\pi})
    \right]
  \right\}.
\end{eqnarray}
This indicates that  
$f^+(q^2)$ and $f^0(q^2)$ scale in the heavy quark limit as
\begin{eqnarray}
  \label{eq:scaling_f+f0}
  f^+(q^2) & \sim & \sqrt{m_B}, \\
  f^0(q^2) & \sim & \frac{1}{\sqrt{m_B}},
\end{eqnarray}
if $v\cdot k_{\pi}$ is kept fixed.

In the soft pion limit $k_{\pi}\rightarrow 0$ and
$m_{\pi}\rightarrow 0$, we obtain simpler relations
\begin{eqnarray}
  \label{eq:f+f0_soft_pion_limit}
  f^+(q^2) & \simeq & 
  \sqrt{m_B} 
  \frac{f_2(v\cdot k_{\pi})}{v\cdot k_{\pi}},
  \\
  f^0(q^2) & \simeq &
  \frac{2}{\sqrt{m_B}}
  \left[
    f_1(v\cdot k_{\pi}) + f_2(v\cdot k_{\pi})
  \right],
\end{eqnarray}
from (\ref{eq:f+_from_f1f2}) and (\ref{eq:f0_from_f1f2}). 
The soft pion theorem implies that the scalar
form factor $f^0(q^2)$ and the $B$ meson leptonic decay
constant $f_B$ are related as 
$f^0(q^2_{max}) = f_B/f_{\pi}$,
which means
\begin{equation}
  \label{eq:soft_pion_theorem}
  f_1(0) + f_2(0) = \frac{f_B\sqrt{m_B}}{2 f_{\pi}}.
\end{equation}
The vector form factor $f^+(q^2)$ may be 
evaluated using the heavy meson chiral lagrangian
approach \cite{Casalbuoni_et_al_97}, in which the $B^*$ pole
contributes through a $B^*B\pi$ coupling.
One obtains the relation
\begin{equation}
  \label{eq:pole_dominance_f2}
  \lim_{v\cdot k_{\pi}\rightarrow 0} f_2(v\cdot k_{\pi}) 
  = g
  \frac{f_{B^*}\sqrt{m_{B^*}}}{2f_{\pi}}
  \frac{v\cdot k_{\pi}}{v\cdot k_{\pi} + \Delta_B},
\end{equation}
where the vector meson decay constant $f_{B^*}$ is defined by
$\langle 0|V^{\mu}|B^*(p)\rangle 
= i f_{B^*}m_{B^*} \epsilon^{\mu}(p)$, and $g$ denotes the
$B^*B\pi$ coupling.
The $B^*$ propagator gives a factor 
$1/(v\cdot k_{\pi} + \Delta_B)$,
in which $\Delta_B = m_{B^*}-m_B$.
Since the hyperfine splitting $\Delta_B\approx 46$~MeV is
much smaller than the `pion' mass, we consider in the lattice
simulation that (\ref{eq:pole_dominance_f2}) depends little on 
on $v\cdot k_{\pi}$.
This behavior of $f_2$ is actually found in our simulation.
Equation (\ref{eq:pole_dominance_f2}) leads to the well-known
vector meson dominance form for the form factor $f^+(q^2)$
\begin{equation}
  \label{eq:pole_dominance_f+}
  \lim_{q^2\rightarrow m_B^2} f^+(q^2) =
  \frac{f_{B^*}}{f_{\pi}}
  \frac{g}{1-q^2/m_{B^*}^2},
\end{equation}
which is also reproduced in our calculation.


%% file: 3.Lattice_NRQCD.tex
\section{Lattice NRQCD}
\label{sec:Lattice_NRQCD}

We use the NRQCD formalism defined on the lattice
\cite{NRQCD} to treat the heavy $b$ quark without large
discretization errors increasing as a power of $aM$.
NRQCD is designed to approximate non-relativistic motion
of heavy quark inside hadrons, and is expressed as a
systematic expansion in some small parameter depending on
the hadron considered.
For the heavy-light meson system such as the $B$ meson, the
expansion parameter is given by $\Lambda_{QCD}/M$, with
$\Lambda_{QCD}$ the typical momentum scale of QCD $\sim$
300--500 MeV.
At the next-to-leading order in $\Lambda_{QCD}/M$, the
lagrangian in the continuum Euclidean space-time is written
as 
\begin{equation}
  \label{eq:NRQCD_continuum}
  {\cal L}_{NRQCD}^{cont} =
  Q^{\dagger} \left[
    D_0 + \frac{\boldvec{D}^2}{2M}
    + g \frac{\boldvec{\sigma}\cdot\boldvec{B}}{2M}
  \right]
  Q,
\end{equation}
for heavy quark field $Q$ represented by a two-component
non-relativistic spinor. 
The derivatives $D_0$ and $\boldvec{D}$ are temporal and
spatial covariant derivatives respectively.
The leading order term $D_0$ represents a heavy quark as a
static color source.
The leading correction of order $\Lambda_{QCD}/M$ comes from
$\boldvec{D}^2/2M$, which gives a non-relativistic kinetic
term of heavy quark.
Another contribution of order $\Lambda_{QCD}/M$ is the
spin-(chromo)magnetic interaction
$\boldvec{\sigma}\cdot\boldvec{B}/2M$, 
where $\boldvec{B}$ denotes a chromomagnetic field strength.
In the usual HQET approach, only the leading terms are present
in the effective lagrangian and corrections of order
$\Lambda_{QCD}/M$ is incorporated when one evaluates a
matrix element $\langle{\cal O}\rangle$ of some operator
${\cal O}$ by including terms such as 
$\langle T {\cal O}
 \int d^4x\,Q^{\dagger}(\boldvec{D}^2/2M)Q \rangle$.
In contrast, in the NRQCD approach we include the correction
terms in the lagrangian (\ref{eq:NRQCD_continuum}) and
evaluate the matrix elements with the heavy quark propagator
including the effect of order $\Lambda_{QCD}/M$.

An important limitation of the NRQCD lagrangian
(\ref{eq:NRQCD_continuum}) is that the heavy quark expansion
is made in the rest frame of a heavy quark.
Since the expansion parameter is $p/M$, where $p$ is a
typical spatial momentum of heavy quark, the lagrangian is
valid only in the region where the heavy quark does not have
momentum greater than $O(\Lambda_{QCD})$.
Therefore, in the study of the heavy-to-light decay, the
momentum of initial $B$ meson must be small enough.
Although it is possible to construct the action expanded
around a finite heavy quark velocity, the heavy quark
velocity is renormalized by radiative correction since the
lattice violates Lorentz symmetry
\cite{Hashimoto_Matsufuru_96,Sloan_lat97}, which gives rise to an
additional important systematic corrections.
We, therefore, do not take this strategy and consider the
discretization of the Lagrangian (\ref{eq:NRQCD_continuum}).

The lattice NRQCD action we use in this work is
\begin{equation}
  \label{eq:NRQCD_lattice}
  S_{NRQCD}= 
  \sum_{x,y} Q^{\dagger}(x)( \delta_{x,y} - K_{Q}(x,y) ) Q(y) +
  \sum_{x,y} \chi^{\dagger}(x)( \delta_{x,y} - K_{\chi}(x.y) ) \chi(y).
\end{equation}
In addition to the non-relativistic heavy quark field $Q$, we
write the term for the anti-particle field $\chi$ for completeness.
The kernels to describe the time evolution of heavy quark are
given by
\begin{eqnarray}
  \label{eq:evolution_kernel_Q}
  K_{Q}(x,y) 
  & = &
  \left[ 
    \left( 1-\frac{a H_{0}}{2 n} \right)^{n}
    \left( 1-\frac{a \delta H}{2} \right)
    \delta^{(-)}_{4}{U^{\dagger}_{4}}
    \left( 1-\frac{a \delta H}{2} \right)
    \left( 1-\frac{a H_{0}}{2 n} \right)^{n}
  \right](x,y), \\
  \label{eq:evolution_kernel_chi}
  K_{\chi}(x,y) 
  & = &
  \left[ 
    \left( 1-\frac{a H_{0}}{2 n} \right)^{n}
    \left( 1-\frac{a \delta H}{2} \right)
    \delta^{(+)}_{4}{U_{4}}
    \left( 1-\frac{a \delta H}{2} \right)
    \left( 1-\frac{a H_{0}}{2 n} \right)^{n}
  \right](x,y),
\end{eqnarray}
where $n$ denotes a stabilization parameter introduced in
order to remove an instability arising from unphysical
momentum modes in the evolution equation \cite{NRQCD}.
The operator $\delta^{(\pm)}_4$ is defined as 
$\delta^{(\pm)}_4(x,y) \equiv \delta_{x_4\pm 1,y_4}
 \delta_{\boldvec{x},\boldvec{y}}$,
and $H_{0}$ and $\delta H$ are lattice Hamiltonians defined
by 
\begin{eqnarray}
  \label{eq:kinetic_term}
  H_{0}      
  & \equiv &
  -\frac{\boldvec{\Delta}^{(2)}}{2 aM_{0}},
  \\
  \label{eq:spin-magnetic}
  \delta{H}
  & \equiv &
  -c_{B} \frac{g}{2aM_{0}} \boldvec{\sigma}\cdot\boldvec{B},
\end{eqnarray}
where 
$\boldvec{\Delta}^{(2)}$ 
$\equiv$ 
$\sum_{i=1}^{3}\Delta^{(2)}_i$ 
is a Laplacian defined on the lattice through
$\Delta^{(2)}_{i}$, the second symmetric covariant
differentiation operator in the spatial direction $i$.
In (\ref{eq:spin-magnetic}) the chromomagnetic field
$\boldvec{B}$ is the usual clover-leaf type lattice field
strength \cite{NRQCD}.
In these definitions, the lattice operators
$\Delta^{(2)}_{i}$ and $\boldvec{B}$ are dimensionless,
\textit{i.e.} appropriate powers of $a$ are understood.
The space-time indices $x$ and $y$ are implicit in these
expressions. 
The bare heavy quark mass $M_0$ is distinguished from the
renormalized one $M$.

The lattice action (\ref{eq:NRQCD_lattice}) describes 
continuum NRQCD (\ref{eq:NRQCD_continuum}) in the limit of
vanishing lattice spacing $a$ at tree level.
In the presence of radiative correction, however, power
divergence of form $\alpha_s^n/(aM_0)^m$ with positive integers
$n,m$ can appear.
This is due to the fact that NRQCD is not
renormalizable, and the action should be considered as an
effective theory valid for small $1/(aM_0)$.
This means that the parameters in the lattice action
(\ref{eq:NRQCD_lattice}) should be tuned to
reproduce the same low energy
amplitude as the continuum QCD up to some higher order
corrections. 
One may use perturbation theory to achieve this tuning. 
For example, one-loop calculation of energy shift and mass
renormalization was carried out for lattice NRQCD 
by Davies and Thacker \cite{Davies_Thacker_92} and by
Morningstar \cite{Morningstar_93} sometime ago, and then by
ourselves
\cite{Hashimoto_et_al_99,JLQCD_fB_00,Hashimoto_et_al_00} for
the above particular form of the NRQCD action.\footnote{
  We note that the evolution kernels
  (\ref{eq:evolution_kernel_Q}) and
  (\ref{eq:evolution_kernel_chi}) are slightly different 
  from the definition used, for example, in
  \cite{Morningstar_93}, where the $(1-aH_0/2n)^n$ terms 
  appear inside of the $(1-a\delta H/2)$ terms.
}
To improve the perturbative expansion we utilize the tadpole
improvement procedure where all the gauge
links in the action (\ref{eq:NRQCD_lattice}) are divided by
its mean field value $u_0$ determined from the plaquette
expectation value as 
$u_0 \equiv (\langle\mathrm{Tr}U_P\rangle/3)^{1/4}$.
This tadpole improvement will give rise to $O(g^2)$ counter
terms in the Feynman rules. 
The one-loop tuning of the coupling constant $c_B$ in front
of the spin-(chromo)magnetic interaction term
(\ref{eq:spin-magnetic}) has not yet been performed.
We, therefore, use the tree level value $c_B$=1 after making
the tadpole improvement.

The relativistic four-component Dirac spinor field $h$ 
is related to the two-component non-relativistic field $Q$
and $\chi$ appearing in the NRQCD action
(\ref{eq:NRQCD_lattice}) via the Foldy-Wouthuysen-Tani (FWT)
transformation  
\begin{equation}
  \label{eq:FWT_transformation}
  h = 
  \left(
    1 -
    \frac{\boldvec{\gamma}\cdot\boldvec{\nabla}}{2aM_0}
  \right)
  \left( 
    \begin{array}{c}
      Q \\ \chi^{\dagger}
    \end{array}
  \right),
\end{equation}
where $\boldvec{\nabla}$ is a symmetric covariant
differentiation operator in a spatial direction.


%% file: 4.Matching.tex
\section{Matching of the heavy-light current}
\label{sec:Matching_of_the_heavy-light_current}

Since we use the lattice NRQCD action of the
previous section, the continuum heavy-light vector current
$\bar{q}\gamma_{\mu}b$ in (\ref{eq:f+f0}) must be written in
terms of the corresponding operator constructed with the
lattice NRQCD heavy quark field $h$.
This matching of the continuum and lattice operators has
been calculated using the one-loop perturbation theory by
Morningstar and Shigemitsu
\cite{Morningstar_Shigemitsu_98,Morningstar_Shigemitsu_99}.
In this section we summarize their results and specify our
notations. 

In the one-loop matching of the continuum operator to the
lattice operators, we have to consider dimension-four
operators in addition to the leading dimension three
operator $\bar{q}\gamma_{\mu}h$, in order to remove the
error of order $\alpha_s\Lambda_{QCD}/M$ and 
$\alpha_s a\Lambda_{QCD}$.
The former is the radiative correction to the FWT
transformation (\ref{eq:FWT_transformation}) and the latter
appears in the $O(a)$-improvement of the lattice discretized
operator. 
Thus the following operators are involved in the
calculation.
\begin{eqnarray}
  \label{eq:V_4^(0)}
  V_4^{(0)} & = & \bar{q} \gamma_4 h,
  \\
  \label{eq:V_4^(1)}
  V_4^{(1)} & = & 
  -\frac{1}{2aM_0} \bar{q} \gamma_4
  \boldvec{\gamma}\cdot\boldvec{\nabla} h,
  \\
  \label{eq:V_4^(2)}
  V_4^{(2)} & = &
  -\frac{1}{2aM_0} \bar{q} 
  \boldvec{\gamma}\cdot\backvec{\boldvec{\nabla}} h,
  \\
  \label{eq:V_k^(0)}
  V_k^{(0)} & = & \bar{q} \gamma_k h,
  \\
  \label{eq:V_k^(1)}
  V_k^{(1)} & = & 
  -\frac{1}{2aM_0} \bar{q} \gamma_k
  \boldvec{\gamma}\cdot\boldvec{\nabla} h,
  \\
  \label{eq:V_k^(2)}
  V_k^{(2)} & = &
  -\frac{1}{2aM_0} \bar{q} 
  \boldvec{\gamma}\cdot\backvec{\boldvec{\nabla}} 
  \gamma_4 \gamma_k h,
  \\
  \label{eq:V_k^(3)}
  V_k^{(3)} & = &
  -\frac{1}{2aM_0} \bar{q}
  \gamma_4 \nabla_k h,
  \\
  \label{eq:V_k^(4)}
  V_k^{(4)} & = &
  \frac{1}{2aM_0} \bar{q}
  \backvec{\nabla}_k \gamma_4 h.
\end{eqnarray}
The heavy quark field $h$ is obtained from the two-component
field $Q$ through the FWT transformation
(\ref{eq:FWT_transformation}).\footnote{
  In the definition used in \cite{Morningstar_Shigemitsu_99}
  the heavy quark field before the FWT transformation
  $(Q\;\;0)^T$
  appears in the definition of operators.
  Matching coefficients for $V_4^{(1)}$ and $V_k^{(1)}$ must
  be converted when we use the above definition.
}
For the light quark $q$ we employ the $O(a)$-improved Wilson
fermion \cite{Sheikholeslami_Wohlert_85}.

The one-loop matching is given by
\begin{eqnarray}
  \label{eq:matching_V4}
  V_4^{cont} & = &
  \left(
    1 + \alpha_s
    \left[
      \frac{1}{\pi}\ln(aM_0) + \rho_{V_4}^{(0)}
    \right]
  \right) V_4^{(0)}
  + \alpha_s\rho_{V_4}^{(1)} V_4^{(1)}
  + \alpha_s\rho_{V_4}^{(2)} V_4^{(2)},
  \\
  \label{eq:matching_Vk}
  V_k^{cont} & = &
  \left(
    1 + \alpha_s
    \left[
      \frac{1}{\pi}\ln(aM_0) + \rho_{V_k}^{(0)}
    \right]
  \right) V_k^{(0)}
  + \alpha_s\rho_{V_k}^{(1)} V_k^{(1)}
  \nonumber\\
  & &
  + \alpha_s
  \left[
    \frac{4}{\pi}\ln(aM_0) + \rho_{V_k}^{(2)}
  \right] V_k^{(2)}
  + \alpha_s\rho_{V_k}^{(3)} V_k^{(3)}
  + \alpha_s
  \left[
    -\frac{4}{3\pi}\ln(aM_0) + \rho_{V_k}^{(4)}
  \right] V_k^{(4)},
\end{eqnarray}
and the numerical coefficients $\rho_{V_4}^{(i)}$ and
$\rho_{V_4}^{(k)}$  are summarized in Tables~\ref{tab:V4}
and \ref{tab:Vk} for several values of $aM_0$.

As we mentioned earlier, the NRQCD action employed in this
work is slightly different from that of Morningstar and
Shigemitsu \cite{Morningstar_Shigemitsu_99}.
We have therefore independently calculated the wave function
renormalization and the vertex correction for the temporal
component $V_4$, and found that the difference of the finite 
constants $\rho$'s between the
two actions is small, \textit{e.g.}, $\sim$ 4--9\% for the vertex correction.
Therefore, for the spatial vector current, for which the
one-loop calculation with our action is missing, we adopt
the coefficients of \cite{Morningstar_Shigemitsu_99}
assuming that the error is negligible.
In Table~\ref{tab:V4} the results of our calculation for
$\rho_{V_4}^{(i)}$ are listed, while results for
$\rho_{V_k}^{(i)}$ in \cite{Morningstar_Shigemitsu_99} are
interpolated in $aM_0$ and given in Table~\ref{tab:Vk} 
for our parameter values.


%% file: 5.Lattice_calculation.tex
\section{Lattice calculation}
\label{sec:Lattice_calculation}

\subsection{Lattice setup}
Our quenched lattice calculation is carried out on a
16$^3\times$48 lattice at $\beta$~=~5.9 with the standard 
plaquette action for gluons.
The inverse lattice spacing $1/a$ determined from the string
tension is $a^{-1}=1.64$ GeV. The scaling violation has been
found to be small for our choice of the heavy and light
quark actions over $1/a \simeq 1 - 2.5$ GeV 
in the heavy-light decay constant 
\cite{JLQCD_fB_00}.

The parameters we choose for the heavy and light quarks are a
subset of those simulated in \cite{JLQCD_fB_00}.
We take four values of the bare mass $aM_0$, 1.3, 2.1, 3.0
and 5.0 for the heavy quark, over a range of the physical
heavy quark mass between 2 and 8 GeV.
The stabilization parameter $n$ is set to 3 (for $aM_0$=1.3
and 2.1) or 2 (for $aM_0$=3.0 and 5.0) so
as to satisfy the stability condition $n>3/(aM_0)$. 
We use the $O(a)-$improved Wilson action for the light quark 
with the clover coefficient $c_{\mathrm{sw}}$=1.580, which
is evaluated at one-loop with the tadpole improvement.
Four values 
0.13630, 0.13711, 0.13769 and 0.13816 are chosen 
for the hopping parameters in our simulation, 
where the critical value $\kappa_c$ is 0.13901.

We accumulate 2150 quenched configurations to reduce the
statistical error for matrix elements with finite spatial
momenta. Each configuration is separated by 1000 pseudo-heat-bath 
sweeps after 10 000 sweeps for thermalization and fixed to 
the Coulomb gauge.
As we will see, even with this large number of statistics,
signals for heaviest heavy quark or lightest light quark
are not clean enough to extract the ground state. 

\subsection{Correlators}
The form factors are extracted from measurements of
three-point correlators
\begin{equation}
  \label{eq:three-point}
  C^{\pi^S V^{(i)}_{\mu} B^S}
  (t_{\pi},t,t_B; \boldvec{q},\boldvec{p}_B)
  =
  \sum_{\boldvec{x},\boldvec{y}}
  e^{-i\boldvec{q}\cdot\boldvec{x}}
  e^{i\boldvec{p}_B\cdot\boldvec{y}}
  \langle 
  {\cal O}_{\pi}^S(t_{\pi},\boldvec{0})
  V_{\mu}^{(i)}(t,\boldvec{x})
  {\cal O}_B^{S\dagger}(t_B,\boldvec{y})
  \rangle,
\end{equation}
of the vector currents
(\ref{eq:V_4^(0)})--(\ref{eq:V_k^(4)}) 
with the initial $B$ meson and daughter pion interpolating
fields ${\cal O}_B^S$ and ${\cal O}_{\pi}^S$, respectively.
The interpolating fields are defined by
\begin{eqnarray}
  {\cal O}_{\pi}^L(t,\boldvec{x})
  & = &
  \bar{q}(t,\boldvec{x}) \gamma_5 q(t,\boldvec{x}),
  \\
  {\cal O}_{\pi}^S(t,\boldvec{x})
  & = &
  \left[
    \sum_{\boldvec{r}} \phi_l(|\boldvec{r}|)
    \bar{q}(t,\boldvec{x}+\boldvec{r})
  \right]
  \gamma_5
  \left[
    \sum_{\boldvec{r}'} \phi_l(|\boldvec{r}'|)
    q(t,\boldvec{x}+\boldvec{r}')
  \right],
  \\
  {\cal O}_B^L(t,\boldvec{x})
  & = &
  \bar{q}(t,\boldvec{x}) \gamma_5 h(t,\boldvec{x}),
  \\
  {\cal O}_B^S(t,\boldvec{x})
  & = &
  \bar{q}(t,\boldvec{x})
  \gamma_5
  \left[
    \sum_{\boldvec{r}} \phi_h(|\boldvec{r}|)
    h(t,\boldvec{x}+\boldvec{r})
  \right],
\end{eqnarray}
where the operators with superscript $L$ represent a local
field, while the smeared operators defined on the Coulomb
gauge fixed configurations are labeled with $S$.
The smearing functions $\phi_l(r)$ and $\phi_h(r)$ are
parameterized by
$\phi_l(r)=\exp(-a_l r^{b_l})$ 
and 
$\phi_h(r)=\exp(-a_h r^{b_h})$, 
with parameters $a_{l,h}$ and $b_{l,h}$ determined from a
measurement of light-light and heavy-light meson wave
functions \cite{JLQCD_fB_00}.
The wave function of the light-light meson $\phi_l(r)$ is
almost independent of the light quark mass, and we use
$(a_l,b_l)$ = (0.27,1.13).
The wave function describing the spread of heavy-light meson
$\phi_h(r)$, on the other hand, depends significantly on the
heavy quark mass, \textit{i.e.} $b_h$ becomes larger as
heavy quark mass increases.
The numerical values of $(a_h,b_h)$ are given in
Table~\ref{tab:smear_parameters}. 

The $B$ meson interpolating field ${\cal O}_B^S$ is fixed at
the time slice $t_B$=24.
The light quark propagator corresponding to the spectator
quark is solved for a smeared source at $t_B$, and the
source method is used at the time slice $t_{\pi}$=0 to obtain
the daughter light quark propagator with momentum insertion
$-\boldvec{q}$.
The heavy-light current is then constructed at $t$, which is
in a region $t_{\pi} < t < t_B$, with the daughter light
quark propagating from $t_{\pi}$ and a heavy anti-quark
evolving back from $t_B$. 
Another momentum $\boldvec{p}_B$ is inserted at $t$.
With this combination of momenta, the initial $B$ meson has
momentum $\boldvec{p}_B$ and the final pion travels with 
momentum $\boldvec{k}_{\pi}=\boldvec{p}_B-\boldvec{q}$,
since the fixed source at $t_B$ emits a heavy-light meson
with any momentum.
The momentum combinations measured in our simulation is
summarized in Table~\ref{tab:momenta}.
Since the statistical noise grows exponentially as
$\exp((E(p^2)-E(0))t)$ for the finite momentum state with
energy $E(p^2)$, the spatial momentum one can measure 
with reasonable signal is rather limited.
In fact, even in our high statistics data, the maximum
momentum we could take is $(1,0,0)$ in unit of $2\pi/La$ as 
we shall discuss in the following sections.

The three-point function (\ref{eq:three-point}) is dominated
by the ground state contribution for large enough separation
of operators $t_{\pi} \ll t \ll t_B$
\begin{eqnarray}
  \label{eq:three-point_asymptotic}
  C^{\pi^S V^{(i)}_{\mu} B^S}
  (t_{\pi},t,t_B; \boldvec{q},\boldvec{p}_B)
  & \rightarrow &
  \frac{Z_{\pi}^S(\boldvec{k}_{\pi})}{2E_{\pi}(\boldvec{k}_{\pi})}
  \frac{Z_B^S(\boldvec{p}_B)}{2E_B(\boldvec{p}_B)}
  \,
  \langle\pi(k_{\pi})|V_{\mu}^{(i)}|B(p_B)\rangle,
  \nonumber\\
  & &
  \times
  \exp\left(
    -E_{\pi}(\boldvec{k}_{\pi})(t-t_{\pi})
    -E_{\mathrm{bind}}(\boldvec{p}_B)(t_B-t)
  \right).
\end{eqnarray}
The overlap amplitudes $Z_{\pi}^S(\boldvec{k}_{\pi})$ and 
$Z_B^S(\boldvec{p}_B)$ of the interpolating operators with
the corresponding ground state are evaluated from the
two-point correlators defined by
\begin{eqnarray}
  \label{eq:C_piSpiS}
  C^{\pi^S \pi^S}(t_{\pi},t; \boldvec{k}_{\pi})
  & = &
  \sum_{\boldvec{x}}
  e^{i\boldvec{k}_{\pi}\cdot\boldvec{x}}
  \langle
  {\cal O}_{\pi}^S(t_\pi,\boldvec{0})
  {\cal O}_{\pi}^{S\dagger}(t,\boldvec{x})
  \rangle
  \rightarrow 
  \frac{Z_{\pi}^S(\boldvec{k}_{\pi})^2}{2E_{\pi}(\boldvec{k}_{\pi})}
  \,e^{-E_{\pi}(\boldvec{k}_{\pi})(t-t_{\pi})},
  \\
  \label{eq:C_piSpiL}
  C^{\pi^S \pi^L}(t_{\pi},t; \boldvec{k}_{\pi})
  & = &
  \sum_{\boldvec{x}}
  e^{i\boldvec{k}_{\pi}\cdot\boldvec{x}}
  \langle
  {\cal O}_{\pi}^S(t_\pi,\boldvec{0})
  {\cal O}_{\pi}^{L\dagger}(t,\boldvec{x})
  \rangle
  \rightarrow 
  \frac{
    Z_{\pi}^S(\boldvec{k}_{\pi})
    Z_{\pi}^L(\boldvec{k}_{\pi})}{2E_{\pi}(\boldvec{k}_{\pi})}
  \,e^{-E_{\pi}(\boldvec{k}_{\pi})(t-t_{\pi})},
  \\
  \label{eq:C_BSBS}
  C^{B^S B^S}(t,t_B; \boldvec{p}_B)
  & = &
  \sum_{\boldvec{x}}
  e^{-i\boldvec{p}_B\cdot\boldvec{x}}
  \langle
  {\cal O}_B^S(t,\boldvec{x})
  {\cal O}_B^{S\dagger}(t_B,\boldvec{0})
  \rangle
  \rightarrow 
  \frac{Z_B^S(\boldvec{p}_B)^2}{2E_B(\boldvec{p}_B)}
  \,e^{-E_{\mathrm{bind}}(\boldvec{p}_B)(t_B-t)},
  \\
  \label{eq:C_BLBS}
  C^{B^L B^S}(t,t_B; \boldvec{p}_B)
  & = &
  \sum_{\boldvec{x}}
  e^{-i\boldvec{p}_B\cdot\boldvec{x}}
  \langle
  {\cal O}_B^L(t,\boldvec{x})
  {\cal O}_B^{S\dagger}(t_B,\boldvec{0})
  \rangle
  \rightarrow
  \frac{
    Z_B^S(\boldvec{p}_B) Z_B^L(\boldvec{p}_B)
     }{2E_B(\boldvec{p}_B)}
  \,e^{-E_{\mathrm{bind}}(\boldvec{p}_B)(t_B-t)}.
\end{eqnarray}
The ground state energy of the heavy-light meson
$E_{\mathrm{bind}}(\boldvec{p}_B)$ represents a ``binding
energy'', as the bare heavy quark mass is subtracted in the
NRQCD formalism.
In the state normalization in
(\ref{eq:three-point_asymptotic}) and in
(\ref{eq:C_BSBS})--(\ref{eq:C_BLBS}), on the other hand,
the heavy-light meson energy $E_B(\boldvec{p}_B)$ including
the bare heavy quark mass enters in the denominator. 

In practice, we calculate a ratio 
$R^{V_{\mu}^{(i)}}(t,\boldvec{k}_{\pi},\boldvec{p}_B)$
of the three-point and the two-point functions
\begin{equation}
  \label{eq:R}
  R^{V_{\mu}^{(i)}}(t;\boldvec{k}_{\pi},\boldvec{p}_B)
  = 
  \frac{
    C^{\pi^S V_{\mu}^{(i)} B^S}
    (t_{\pi},t,t_B; \boldvec{q},\boldvec{p}_B)
    }{
    C^{\pi^S \pi^L}(t_{\pi},t; \boldvec{k}_{\pi})
    C^{B^L B^S}(t,t_B; \boldvec{p}_B)
    }
  \rightarrow
  \frac{
    \langle\pi(k_{\pi})|V_{\mu}^{(i)}|B(p_B)\rangle
    }{
    Z_{\pi}^L(\boldvec{k}_{\pi})
    Z_B^L(\boldvec{p}_B)},
\end{equation}
which becomes constant in the asymptotic limit.
The overlap amplitudes with the smeared interpolating
fields $Z_{\pi}^S(\boldvec{k}_{\pi})$ and
$Z_B^S(\boldvec{p}_B)$ cancel between the numerator and 
the denominator. 
Typical examples of the ratio
$R^{V_{\mu}^{(i)}}(t;\boldvec{k}_{\pi},\boldvec{p}_B)$ is
plotted in Figure~\ref{fig:R}, in which the data at
$\kappa$=0.13711 and $aM_0$=3.0 are shown for five choices
of the momentum combinations.
For all these plots we find clear plateau in the large $t$
region, where the current is closer to the $B$ meson
interpolating field than to the pion.
This is due to the fact that the smearing with the measured
wave function works better for heavy-light than for
light-light meson, because the heavy-light system is less
relativistic and the description with the wave function is
more appropriate.
The fit result is indicated by horizontal lines.

The data become noisier for lighter light quark masses with a fixed
heavy quark mass, or for heavier heavy quark mass with fixed light
quark mass.
As a result, we are not able to extract
signals for our lightest light quark $\kappa$=0.13816,
except for a few cases when the daughter pion does not have
finite spatial momentum.
We also note that we carried out simulations for one
additional heavy quark mass $aM_0$=10.0.
We found, however, that the signal is intolerably noisy,
so that we do not use those data in our analysis.

\subsection{Matrix elements}
In order to obtain the matrix element
$\langle\pi(k_{\pi})|V_{\mu}^{(i)}|B(p_B)\rangle$
>from (\ref{eq:R}), we have to eliminate 
$Z_{\pi}^L(\boldvec{k}_{\pi}) Z_B^L(\boldvec{p}_B)$
in the denominator.
For this purpose we fit the smeared-smeared and
smeared-local two-point functions with a single exponential 
as in
(\ref{eq:C_piSpiS}) and (\ref{eq:C_piSpiL}) for the
extraction of
$Z_{\pi}^L(\boldvec{k}_{\pi})/\sqrt{E_{\pi}(\boldvec{k}_{\pi})}$,
and 
(\ref{eq:C_BSBS}) and (\ref{eq:C_BLBS}) for
$Z_B^L(\boldvec{p}_B)/\sqrt{E_B(\boldvec{p}_B)}$.
We then obtain a combination
\begin{equation}
  \label{eq:V_hat}
  \hat{V}_{\mu}^{(i)}(\boldvec{k}_{\pi},\boldvec{p}_B)
  \equiv
  \frac{
    \langle\pi(k_{\pi})|V_{\mu}^{(i)}|B(p_B)\rangle
    }{
    \sqrt{E_{\pi}(\boldvec{k}_{\pi}) E_B(\boldvec{p}_B)}
    }.
\end{equation}
Numerical results are listed in
Tables~\ref{tab:V.k1}--\ref{tab:V.k4} for each light and
heavy quark mass.
The first column denotes the momentum configuration as shown
in Table~\ref{tab:momenta}.


%% file: 6.Results.tex
\section{Results for the form factors}
\label{sec:Results_for_the_form_factors}

\subsection{Energy-momentum dispersion relations}
In order to extract the form factors from the matrix
elements (\ref{eq:V_hat}), we have to determine the meson
energy of initial and final states for given spatial
momenta. 
It may be obtained either by assuming a continuum dispersion
relation or by actually measuring the meson energy with the
given momenta.

For the pion, which is relativistic, the continuum
dispersion relation is written as
\begin{equation}
  \label{eq:dispersion_pi}
  E_{\pi}(\boldvec{k}_{\pi})^2
  = M_{\pi}^2 + \boldvec{k}_{\pi}^2.
\end{equation}

The measured values of $(aE_{\pi}(\boldvec{k}_{\pi}))^2$ for 
momenta $\boldvec{k}_{\pi}$ = (1,0,0) and (1,1,0), in unit
of $2\pi/La$, are given in Table~\ref{tab:pion_energy} and
also plotted in Figure~\ref{fig:dispersion_pi}
for each light quark mass we calculated.
We find a nice agreement with the expectation
(\ref{eq:dispersion_pi}). 
The relation (\ref{eq:dispersion_pi}) may be modified on the
lattice due to lattice artifacts; a possible form is
given by replacing $a\boldvec{k}_{\pi}$ with
$\sin(a\boldvec{k}_{\pi})$, which satisfies the periodic
boundary condition.
The magnitude of such an effect is not significant, though, since the momentum
considered is small enough and the difference between
$a\boldvec{k}_{\pi}$ and $\sin(a\boldvec{k}_{\pi})$ is less
than 3\%. 

The dispersion relation for the heavy-light meson is well
described by the non-relativistic form
\begin{equation}
  \label{eq:dispersion_B}
  E_{\mathrm{bin}}(\boldvec{p}_B)
  = E_{\mathrm{bin}}(\boldvec{0}) +
  \frac{\boldvec{p}_B^2}{2M_B},
\end{equation}
in which the meson mass $M_B$ appears in the kinetic energy
term.\footnote{
  Here we use a capital symbol $M_B$ to represent the
  generic heavy-light meson mass we deal with on the
  lattice, whereas keeping $m_B$ to denote the physical $B$
  meson mass.}
In NRQCD, the heavy-light meson mass is written in terms of
the bare mass $aM_0$ and the binding energy
$aE_{\mathrm{bin}}(\boldvec{0})$ as
\begin{equation}
  \label{eq:heavy-light_meson_mass}
  aM_B = Z_m aM_0 - aE_0 + aE_{\mathrm{bin}}(\boldvec{0}),
\end{equation}
where $aE_0$ is an energy shift and $Z_m$ is a mass
renormalization factor. 
Both factors are calculated at the one-loop level
\cite{Davies_Thacker_92,Morningstar_93,JLQCD_fB_00},
\begin{eqnarray}
  \label{eq:E_0}
  aE_0 & = & \alpha_s A, \\
  \label{eq:Z_m}
  Z_m & = & 1 + \alpha_s B,
\end{eqnarray}
and the numerical coefficients $A$ and $B$ are given in
Table~I of \cite{JLQCD_fB_00}.
The heavy-light meson mass evaluated with
(\ref{eq:heavy-light_meson_mass}) using the $V$-scheme
coupling $\alpha_V(q^*)$ \cite{Lepage_Mackenzie_93} at
$q^*=1/a$ is listed in Table~\ref{tab:M_B}.
Since the one-loop correction partially cancels between 
$Z_m aM_0$ and $aE_0$, the uncertainty due to the choice of $q^*$ is
small, \textit{i.e.} at most 3\% for $aM_0$=1.3 and even
smaller for larger $aM_0$.

In Figure~\ref{fig:dispersion_B}, a comparison is made of
our simulation data with the form of Eq.~(\ref{eq:dispersion_B})
in which the value of $M_B$ evaluated according to
Eq.~(\ref{eq:heavy-light_meson_mass}) is substituted.
We find a good agreement except for the data at
$\kappa$=0.13630.
Even in the worst case, the disagreement does not exceed
1\%.
Therefore, we employ the dispersion relation
(\ref{eq:dispersion_B}) with the perturbatively estimated
meson mass $aM_B$ in the following analysis of the form
factors, rather than using the measured binding energy,
which has significant statistical errors and complicates our
analysis. 
The same strategy is taken for the pion energy, namely we
use the relation (\ref{eq:dispersion_pi}) with the measured value for 
$aM_{\pi}$.

\subsection{Form factor extraction}
The continuum matrix element is obtained from
$\hat{V}_{\mu}^{(i)}(\boldvec{k}_{\pi},\boldvec{p}_B)$
defined in (\ref{eq:V_hat}) using the matching formula of
the vector current
(\ref{eq:matching_V4})--(\ref{eq:matching_Vk}) as 
\begin{eqnarray}
  \label{eq:V4_hat_cont}
  \hat{V}_4^{cont}(\boldvec{k}_{\pi},\boldvec{p}_B)
  & = &
  \left(
    1+\alpha_s\left[
      \frac{1}{\pi}\ln(aM_0)+\rho_{V_4}^{(0)}
    \right]
  \right) 
  \hat{V}_4^{(0)}(\boldvec{k}_{\pi},\boldvec{p}_B)
  + \alpha_s\rho_{V_4}^{(1)} 
  \hat{V}_4^{(1)}(\boldvec{k}_{\pi},\boldvec{p}_B)
  \nonumber\\
  & &
  + \alpha_s\rho_{V_4}^{(2)} 
  \hat{V}_4^{(2)}(\boldvec{k}_{\pi},\boldvec{p}_B),
  \\
  \label{eq:Vk_hat_cont}
  \hat{V}_k^{cont}(\boldvec{k}_{\pi},\boldvec{p}_B)
  & = &
  \left(
    1+\alpha_s\left[
      \frac{1}{\pi}\ln(aM_0)+\rho_{V_k}^{(0)}
    \right]
  \right) 
  \hat{V}_k^{(0)}(\boldvec{k}_{\pi},\boldvec{p}_B)
  + \alpha_s\rho_{V_k}^{(1)} 
  \hat{V}_k^{(1)}(\boldvec{k}_{\pi},\boldvec{p}_B)
  \nonumber\\
  & &
  + \alpha_s\left[
    \frac{4}{\pi}\ln(aM_0)+\rho_{V_k}^{(2)} 
  \right]
  \hat{V}_k^{(2)}(\boldvec{k}_{\pi},\boldvec{p}_B)
  + \alpha_s\rho_{V_k}^{(3)} 
  \hat{V}_k^{(3)}(\boldvec{k}_{\pi},\boldvec{p}_B)
  \nonumber\\
  & &
  + \alpha_s\left[
    -\frac{4}{3\pi}\ln(aM_0)+\rho_{V_k}^{(4)} 
  \right]
  \hat{V}_k^{(4)}(\boldvec{k}_{\pi},\boldvec{p}_B).
\end{eqnarray}
We use  the $V$-scheme
coupling $\alpha_V(q^*)$ for the coupling constant $\alpha_s$.
Since the scale $q^*$  which dominates the lattice one-loop integral 
 is not yet known, we examine the uncertainty in
the scale setting by calculating the form factors at $q^* = 1/a$ and at
$\pi/a$.
We use the difference in the results, which is the two-loop
effect of $O(\alpha_s^2)$, as an
estimate of higher order perturbative errors.
The numerical value of the coupling is $\alpha_V(1/a)$=0.270
and $\alpha_V(\pi/a)$=0.164 at $\beta$=5.9 in the quenched
approximation. 

From the definition of 
$f_1(v\cdot k_{\pi})$ and $f_2(v\cdot k_{\pi})$ given in
(\ref{eq:f1f2}), we obtain the following formula for
the form factors
\begin{eqnarray}
  \label{eq:f1+f2_from_V_hat}
  f_1(v\cdot k_{\pi}) + f_2(v\cdot k_{\pi})
  & = &
  \sum_{\mu}
  v^{\mu}
  \left[
    \sqrt{
      \frac{
        E_{\pi}(\boldvec{k}_{\pi}) E_B(\boldvec{p}_B)
        }{4 M_B}
      }
    \hat{V}_{\mu}^{cont}(\boldvec{k}_{\pi},\boldvec{p}_B)
  \right],
  \\
  \label{eq:f2_from_V_hat}
  f_2(v\cdot k_{\pi})
  \left[
    1 - \frac{M_{\pi}^2}{(v\cdot k_{\pi})^2}
  \right]
  & = &
  \sum_{\mu}
  \left(
    v^{\mu} - \frac{k_{\pi}^{\mu}}{(v\cdot k_{\pi})}
  \right)
  \left[
    \sqrt{
      \frac{
        E_{\pi}(\boldvec{k}_{\pi}) E_B(\boldvec{p}_B)
        }{4 M_B}
      }
    \hat{V}_{\mu}^{cont}(\boldvec{k}_{\pi},\boldvec{p}_B)
  \right],
\end{eqnarray}
where 
$v^{\mu} = (E_B(\boldvec{p}_B), \boldvec{p}_B)/M_B$
and
$k_{\pi}^{\mu} 
 = (E_{\pi}(\boldvec{k}_{\pi}), \boldvec{k}_{\pi})$.
By construction, for the initial $B$ meson at rest,
$f_1(v\cdot k_{\pi})+f_2(v\cdot k_{\pi})$ is proportional to
the temporal component
$\hat{V}_4^{cont}(\boldvec{k}_{\pi},\boldvec{p}_B)$,
while $f_2(v\cdot k_{\pi})$ comes from the spatial component 
$\hat{V}_k^{cont}(\boldvec{k}_{\pi},\boldvec{p}_B)$.
Even for a $B$ meson with momentum (1,0,0), the velocity is
small ($\boldvec{p}_B/M_B \simeq$ 0.07--0.2 depending on the
heavy quark mass), and the major effect is from the temporal
or spatial component for 
$f_1(v\cdot k_{\pi})+f_2(v\cdot k_{\pi})$ 
or for $f_2(v\cdot k_{\pi})$, respectively.

An example of the form factors is plotted in
Figure~\ref{fig:f1f2_typical} for $aM_0$=3.0, which is close
to the $b$ quark mass, and $\kappa$=0.13630.
The point of smallest $a v\cdot k_{\pi}$ corresponds to the
zero recoil configuration, \textit{i.e.} the initial and
final particles are at rest so that
$a v\cdot k_{\pi} = aM_{\pi}$.
At that point, only the temporal component
$\hat{V}_4^{cont}(\boldvec{k}_{\pi},\boldvec{p}_B)$ can be
measured while the spatial component vanishes.
The momentum configuration $\boldvec{p}_B$=(1,0,0) and
$\boldvec{k}_{\pi}$=(0,0,0) gives a very similar 
$a v\cdot k_{\pi}$, because of large heavy quark mass and
small spatial velocity.
As a result, the data point almost lies on top of that at zero 
recoil.
We are not able to measure $f_2(v\cdot k_{\pi})$ reliably at
this point, since the value of the spatial component 
$\hat{V}_1^{cont}(\boldvec{k}_{\pi},\boldvec{p}_B)$ is too
small. 
There are four other momentum configurations (see
Table~\ref{tab:momenta}), for which both 
$f_1(v\cdot k_{\pi})+f_2(v\cdot k_{\pi})$ 
and $f_2(v\cdot k_{\pi})$ are measured.
Among them, two momentum configurations sharing the same
$\boldvec{k}_{\pi}$=(1,0,0) and having different
$\boldvec{p}_B$ have almost identical values of 
$a v\cdot k_{\pi}$ for the same reason as above, and cannot be
distinguished from each other in the plot (the middle point
of the three filled data points).

From Figure~\ref{fig:f1f2_typical} we also see
that the effect from choosing
$\alpha_V(1/a)$ (circles) or $\alpha_V(\pi/a)$ (squares) is
small; it is smaller than the statistical error except for
the zero recoil point where the statistical error is minimum.
Therefore in the following analysis we use the data with
$\alpha_V(1/a)$.
In the final results we will include
their difference in the systematic error estimation.

\subsection{Heavy quark mass dependence}
\label{sec:Heavy_quark_mass_dependence}
As we discussed in Section~\ref{sec:HQET_form_factors}, the
heavy quark scaling is manifest for the form factors 
$f_1(v\cdot k_{\pi})$ and $f_2(v\cdot k_{\pi})$.
Namely, $f_{1,2}(v\cdot k_{\pi})$ behaves as a constant at
the leading order of the $1/M$ expansion.
Here we examine the heavy quark mass dependence of 
$f_1(v\cdot k_{\pi})+f_2(v\cdot k_{\pi})$ 
and $f_2(v\cdot k_{\pi})$ explicitly by comparing the
results with different heavy quark masses.

In order to remove the logarithmic dependence on the heavy
quark mass that appears from the matching of the vector
current between the full QCD and lattice NRQCD
(\ref{eq:matching_V4})--(\ref{eq:matching_Vk}), we define
the renormalization group invariant form factors
$\Phi_{1+2}(v\cdot k_{\pi})$ and $\Phi_2(v\cdot k_{\pi})$ as 
\begin{eqnarray}
  \label{eq:Phi_1+2}
  \Phi_{1+2}(v\cdot k_{\pi}) & = &
  \left(
    \frac{\alpha_s(M_B)}{\alpha_s(m_B)}
  \right)^{2/\beta_0}
  \left[
    f_1(v\cdot k_{\pi}) + f_2(v\cdot k_{\pi})
  \right],
  \\
  \label{eq:Phi_2}
  \Phi_2(v\cdot k_{\pi}) & = &
  \left(
    \frac{\alpha_s(M_B)}{\alpha_s(m_B)}
  \right)^{2/\beta_0}
  f_2(v\cdot k_{\pi}),
\end{eqnarray}
where $\beta_0$ denotes the first coefficient of the QCD
beta function.
We note that $M_B$ is the heavy-light meson mass measured on
the lattice for a given $aM_0$, while $m_B$ is the physical
$B$ meson mass.

Figure~\ref{fig:f1f2_heavy_dependence} shows 
$\Phi_{1+2}(v\cdot k_{\pi})$ and $\Phi_2(v\cdot k_{\pi})$
for several values of $aM_0$.
We find that the $1/M$ correction gives only a small effect in the
range of the heavy quark mass we explored which correspond
to 2--8 GeV.
In fact, there is no significant shift in the magnitude of
the form factors by the change of the heavy quark mass.
A small effect can be seen in the value of 
$a v\cdot k_{\pi}$ for a couple of momentum configurations
for which $\boldvec{p}_B\cdot\boldvec{k}_{\pi} \neq 0$.
However, it does not seem to change the global shape of the
form factors.

The small $1/M$ correction we found is of great
phenomenological importance, as it justifies the use of 
heavy quark symmetry to predict the 
$B\rightarrow\pi l\nu$ form factors from that of
$D\rightarrow\pi l\nu$ and $D\rightarrow K l\nu$
\cite{Burdman_Ligeti_Neubert_Nir_94}. 
We discuss this method and possible uncertainties
in Section~\ref{sec:Phenomenological_Implications}.

\subsection{Light quark mass dependence}
\label{sec:Light_quark_mass_dependence}
In order to obtain the physical form factors
we need to extrapolate our result to the  physical light ($u$ and $d$) 
quark mass. For this purpose we examine the light quark mass dependence of the
form factors $f_1(v\cdot k_{\pi})+f_2(v\cdot k_{\pi})$ and
$f_2(v\cdot k_{\pi})$ using the data, which covers a range
0.45--0.80 GeV of the pion mass. 
Unfortunately, the signal is badly contaminated by
statistical noise for the lightest data, so that we are not
able to extract the form factors except for the zero recoil
limit of $f_1(v\cdot k_{\pi})+f_2(v\cdot k_{\pi})$.
For other three $\kappa$ values, the data is fully available
and we mainly use them to see the light quark mass
dependence.

Figure~\ref{fig:f1f2_light_dependence} shows the measured
form factors at four different light quark masses.
The heavy quark mass is fixed at $aM_0$=2.1.
Since the minimum value of $av\cdot k_{\pi}$ is $aM_{\pi}$,
the range of $av\cdot k_{\pi}$ where the data is available
moves to the left hand side as the light quark mass
decreases.
On the other hand, change of the value of the form factors 
$f_1(v\cdot k_{\pi})+f_2(v\cdot k_{\pi})$ and
$f_2(v\cdot k_{\pi})$ 
is not significant if we compare the data for a given
momentum configuration.
For instance, the values of 
$a^{1/2}[f_1(v\cdot k_{\pi})+f_2(v\cdot k_{\pi})]$
stay almost constant around 0.68 
over the range $v\cdot k_{\pi} =$ 0.27 -- 0.49, which correspond to 
the lightest and the heaviest data.
If we look at the change at fixed $v\cdot k_{\pi}$, there is
an apparent downward shift of 
$f_1(v\cdot k_{\pi})+f_2(v\cdot k_{\pi})$.
This is due to a negative slope in $av\cdot k_{\pi}$ in
the data at fixed light quark mass.
On the other hand, for $f_2(v\cdot k_{\pi})$ the light quark
mass dependence is less significant, since the data at fixed
$\kappa$ does not seem to have a non-zero slope.

\subsection{Global fit}
In order to extract the physical form factors, we have to
consider the dependence on three parameters, \textit{i.e.}
the inverse heavy meson mass $1/M_B$, the light quark mass
$m_q$ and the energy release $v\cdot k_{\pi}$.
The heavy quark effective theory together with the chiral
perturbation theory suggests that we can expand the form
factors in powers of $1/M_B$ and $m_q$. On the other hand, there is no
theoretical guide in the functional dependence on 
$v\cdot k_{\pi}$.
Therefore, in fitting the data we use a Taylor expansion
around an arbitrary chosen point 
$v\cdot k_{\pi}=(v\cdot k_{\pi})_0$, which in practice we
take in the middle of the measured range.
Thus we employ the following form to fit the data,
\begin{eqnarray}
  \label{eq:global_fit_f1+2}
  a^{1/2} \Phi_{1+2}(v\cdot k_{\pi})
  & = &
  C_{1+2}^{(000)} + \frac{C_{1+2}^{(100)}}{aM_B}
  + C_{1+2}^{(010)} am_q
  + 
  \left(
    C_{1+2}^{(001)} + C_{1+2}^{(011)} am_q
  \right)
  \left[
    av\cdot k_{\pi} - (av\cdot k_{\pi})_0
  \right]
  \nonumber\\
  & &
  + C_{1+2}^{(002)}
  \left[
    av\cdot k_{\pi} - (av\cdot k_{\pi})_0
  \right]^2,
  \\
  \label{eq:global_fit_f2}
  a^{1/2} \Phi_{2}(v\cdot k_{\pi})
  & = &
  C_{2}^{(000)} + \frac{C_{2}^{(100)}}{aM_B}
  + C_{2}^{(010)} am_q
  + C_{2}^{(001)} 
  \left[
    av\cdot k_{\pi} - (av\cdot k_{\pi})_0
  \right],
\end{eqnarray}
where the superscript $(ijk)$ for the coefficient denotes
the order of expansion in $1/aM_B$, $am_q$ and 
$[av\cdot k_{\pi} - (av\cdot k_{\pi})_0]$, in the given
order. 
The fit results for $(av\cdot k_{\pi})_0$=0.5 is
listed in Table~\ref{tab:global_fit_parameters}.

The choice whether we keep a certain term in
(\ref{eq:global_fit_f1+2})--(\ref{eq:global_fit_f2}) or 
drop it is empirical.
Our experience in the calculations of the heavy-light decay
constant and the $B$ parameters suggests that both the
$1/M_B$ and $am_q$ expansions can be safely truncated at the
first order. This is consistent with an argument of naive
power counting assuming that the relevant mass scale is
around $\Lambda_{QCD}$.
We find that that is indeed the case also for the 
$B\rightarrow\pi l\nu$
form factors as we shall discuss in the following.

In (\ref{eq:global_fit_f1+2})--(\ref{eq:global_fit_f2}) the
$1/M_B$ expansion is truncated at first order, since the
$1/M_B$ correction is not significant as discussed in
Section~\ref{sec:Heavy_quark_mass_dependence} so that there
is no sensitivity to higher order corrections.
Even the first order coefficients $C_{1+2}^{(100)}$ and
$C_2^{(100)}$ are consistent with zero within the
statistical error.
This truncation is also consistent with our NRQCD action,
because we do not include the correction of order $1/M_0^2$
or higher in the action (\ref{eq:NRQCD_lattice}).

The light quark mass dependence of 
$a^{1/2} \Phi_{1+2}(v\cdot k_{\pi})$ is consistent with 
a linear function if we fix $av\cdot k_{\pi}$ at 
$(av\cdot k_{\pi})_0$=0.5, for instance.
Thus we truncate the expansion in $am_q$ at the first order.
We also keep a cross term with the leading 
$[av\cdot k_{\pi} - (av\cdot k_{\pi})_0]$ correction, but
its coefficient $C_{1+2}^{(011)}$ is consistent with zero.
For $a^{1/2} \Phi_2(v\cdot k_{\pi})$ the light quark mass
dependence is not significant as discussed in
Section~\ref{sec:Light_quark_mass_dependence}.
Although we keep the first order correction to be
conservative, its coefficient $C_2^{(010)}$ is almost
consistent with zero. 

As for the functional dependence of the form factors on 
$a v\cdot k_{\pi}$,  
we include the $[av\cdot k_{\pi} - (av\cdot k_{\pi})_0]^2$
term for $a^{1/2} \Phi_{1+2}(v\cdot k_{\pi})$, while
the second order term is neglected for 
$a^{1/2} \Phi_2(v\cdot k_{\pi})$.
The reason is that we find a significant slope in
$a^{1/2} \Phi_{1+2}(v\cdot k_{\pi})$, so that a higher order term
$[av\cdot k_{\pi} - (av\cdot k_{\pi})_0]^2$ is also included
for safety.
The other form factor $a^{1/2} \Phi_2(v\cdot k_{\pi})$
behaves almost like constant, and it is enough to keep the
first order term.

While we introduce several terms for which the
coefficient is not well determined, \textit{i.e.} consistent
with zero, this does not mean our results for the physical
form factors have large uncertainty, as far as we use 
the results for an interpolation in the relevant parameters.
For example, the heavy quark mass we simulate covers the
$b$ quark mass, so that only an interpolation is required.
For the parameter
$[av\cdot k_{\pi} - (av\cdot k_{\pi})_0]$,
we restrict ourselves to consider the region where the data
is available.
Therefore, we can obtain the physical form factors in the
region 0.67 GeV $< v\cdot k_{\pi} <$ 0.96 GeV reliably.
Outside this region, the fit 
(\ref{eq:global_fit_f1+2})--(\ref{eq:global_fit_f2})
appears to introduce a large uncertainty.
For the light quark mass, we have to consider an
extrapolation to the physical limit of $u$ and $d$ quarks.
This increases our statistical error significantly.

The fit results are shown in
Figure~\ref{fig:f1f2_heavy_dependence} (heavy quark mass
dependence) and in Figure~\ref{fig:f1f2_light_dependence}
(light quark mass dependence).
In Figure~\ref{fig:f1f2_light_dependence} we also plot the
limit of physical light quark mass (thick curves), which is
obtained by setting $am_q$ to the physical average up and
down quark masses in 
(\ref{eq:global_fit_f1+2})--(\ref{eq:global_fit_f2}).

The form factors 
$f_1(v\cdot k_{\pi})+f_2(v\cdot k_{\pi})$
and $f_2(v\cdot k_{\pi})$ 
for the physical $B\rightarrow\pi l\nu$ decay is plotted in
Figure~\ref{fig:f1f2_physical}.
The region where the lattice data is interpolated in 
$[av\cdot k_{\pi} - (av\cdot k_{\pi})_0]$ is plotted with
symbols.
Going outside of that region requires an extrapolation, so
that the error shown by dashed curves rapidly grows.

\subsection{Soft pion theorem}
\label{sec:Soft_pion_theorem}
In the soft pion limit, \textit{i.e.} $m_{\pi}$ and
$k_{\pi}\rightarrow 0$, the following relation 
(\ref{eq:soft_pion_theorem}) holds
\begin{equation}
  \label{eq:soft_pion_theorem_f0}
  f^0(q^2_{max}) 
  = \frac{2}{\sqrt{M_B}} \left[ f_1(0)+f_2(0) \right]
  = \frac{f_B}{f_{\pi}}.
\end{equation}
It is an important consistency check to see whether one can
reproduce this relation in the lattice calculation.

In Figure~\ref{fig:soft_pion_limit} we plot the result of
the fit (\ref{eq:global_fit_f1+2}) by an open triangle
and compare it with the lattice calculation of $f_B/f_{\pi}$ (filled
triangle) \cite{JLQCD_fB_00}.
The data is presented at fixed heavy quark mass $aM_0$=3.0. 
We should note that the soft pion limit in
(\ref{eq:global_fit_f1+2}) is far from the region where 
$f_1(v\cdot k_{\pi})+f_2(v\cdot k_{\pi})$ is obtained by
interpolation. 
Therefore, we expect substantial systematic uncertainty in
the fit result. 
In fact, Figure~\ref{fig:f1f2_physical} demonstrates 
that the extrapolation to
$v\cdot k_{\pi}=0$ is not yet very stable.

The soft pion limit can also be achieved along a fixed
momentum configuration.
Namely we may extrapolate the data for each light quark mass
at zero recoil.
In this case, however, the momentum transfer 
$v\cdot k_{\pi} (=M_{\pi})$ changes during the
extrapolation, so that we have to consider a fit with two
terms $am_q$ and $aM_{\pi}$.\footnote{
  As discussed in \cite{UKQCD_lat00}, one should include a
  term which is linear in $aM_{\pi}$ when $v\cdot k_{\pi}$
  (or $q^2$ in the relativistic form) varies during the
  extrapolation of form factors. 
  The fit becomes more stable if one first interpolates to a
  fixed $v\cdot k_{\pi}$ (or $q^2$), and then extrapolate in
  $am_q$.
  Our strategy to employ the global
  fit (\ref{eq:global_fit_f1+2})--(\ref{eq:global_fit_f2})
  is equivalent to this method.
}
Because of the PCAC relation $M_{\pi}^2\propto m_q$, it
means a quadratic fit in $\sqrt{am_q}$.
We plot two extrapolations in
Figure~\ref{fig:soft_pion_limit}:
a linear form in $am_q$ (dashed line) and
a quadratic fit in $\sqrt{am_q}$ (solid curve). 
Although the effect of the term $\sqrt{am_q}$ seems very
small in the data and is only seen at the lightest quark
mass, it raises  the soft pion limit for the quadratic fit. 
The result is consistent with the global fit
(\ref{eq:global_fit_f1+2}). 
Thus we consider that the disagreement of $f^0(q^2_{max})$ with
$f_B/f_{\pi}$, which seemed to be a serious problem by only looking
at the naive linear extrapolation with only data at zero recoil,
is in fact more of a problem in the subtle chiral extrapolation
or in the model uncertainty of momentum extrapolation. 

The UKQCD \cite{UKQCD_lat99,UKQCD_lat00} and APE
\cite{APE_00,APE_lat00} collaborations found in their
study with the relativistic heavy quark action that the soft
pion relation (\ref{eq:soft_pion_theorem_f0}) is satisfied. 
It should be noted, however, that their method of chiral 
extrapolation corresponds to our ``global fit'' method, and
the measured kinematical region is far from the soft
pion limit. 
Therefore the result in the soft pion limit should depend
on how the extrapolation is made.
They employed a pole-like model \cite{Becirevic_Kaidalov_00} for 
their fit function. Thus their results in the soft pion limit 
contain some uncertainty which is not well controlled just
as ours, although the results in the kinematical region
obtained by interpolating the lattice data do not suffer
from such uncertainties. 

Judging from the size of the uncertainties it is too early 
to consider the deviation from the soft pion relation as a serious 
problem. This problem can be studied with much statistically 
significant data with larger number of momentum points and 
light quark masses so that the extrapolation in $v\cdot k_{\pi}$
towards the soft pion limit becomes more stable, which 
is still beyond the scope of this paper.

\subsection{Pole dominance}
\label{sec:Pole_dominance}
In the soft pion limit, the heavy meson effective lagrangian
predicts the $B^*$ pole dominance
(\ref{eq:pole_dominance_f2}), that is 
\begin{equation}
  \lim_{v\cdot k_{\pi}\rightarrow 0} f_2(v\cdot k_{\pi}) 
  = g
  \frac{f_{B^*}\sqrt{m_{B^*}}}{2f_{\pi}}
  \frac{v\cdot k_{\pi}}{v\cdot k_{\pi} + \Delta_B}.
\end{equation}
Since the hyperfine splitting 
$\Delta_B\equiv M_{B^*}-M_B$ is much smaller than the
momentum transfer $v\cdot k_{\pi}$ we measure, we can
approximate its functional form by a constant in our data
region. 
Our data support the constant behavior and give
$g(f_{B^*}\sqrt{a m_{B^*}}/2f_{\pi})$ = 0.35(18)
which reduces to $g$ = 0.30(16).
This agrees with the phenomenological value extracted from
$D^*\rightarrow D\pi$ decay 0.27(6) \cite{Stewart_98},
and also with the recent lattice calculation 
$g$ = 0.42(4)(8) \cite{UKQCD_98}, which is obtained for the
static heavy quark.
The agreement suggests that the $1/M$ correction is small
for the form factors.

\subsection{Systematic Errors}
We now discuss possible sources of systematic errors and
their estimates. 
Since the statistical error, the discretization error of
$O(a^2)$, the perturbative error of $O(\alpha_s^2)$, and
the chiral extrapolation error are large, 
we only consider these dominant sources of errors and
neglect other subleading errors such as 
$O(\alpha_s^2/(aM))$, $O(\alpha_s^2 a \Lambda_{QCD})$, 
$O(\alpha_s \Lambda_{QCD}/M)$ and so on. 

The size of the two-loop order correction is only known
by explicit computation, which is beyond the scope of this paper. 
Instead, we estimate the size of the perturbative error of
$O(\alpha_s^2)$ as half of the difference of values for 
$q^{\ast}=\pi/a$ and $1/a$. The typical sizes are 1.5\% for 
$f_1(v\cdot k_\pi)+f_2(v\cdot k_\pi)$
and 3.5\% for $f_2(v\cdot k_\pi)$.
The reason for the error for $f_2(v\cdot k_\pi)$ being
larger is that the one-loop renormalization coefficient for
heavy-light vector current is larger for the spatial
component than for the temporal one and that the matrix element
of the spatial component gives larger contributions to 
$f_2(v\cdot k_\pi)$ in the small recoil region. 

The discretization errors of $O(a^2\Lambda_{QCD}^2)$ and of
$O(a^2 \boldvec{k}_{\pi}^2)$ are also important. 
The former error is common to most lattice simulations 
using the $O(a)$-improved actions,
and through an order counting we estimate it to be 3\% at
$\beta=5.9$ assuming that the typical momentum scale
$\Lambda_{QCD}$ is around 300~MeV. 
The latter is specific to the present work since the error
due to nonzero recoil momenta appears only in the study of form
factors. 
As the pion momentum treated in our calculation is at most
$2\pi/L$ ($L$=16) in the lattice unit, we estimate this
error to be about 16\% using the order estimation.

The error in the chiral extrapolation is another major
source of the systematic error. 
Since we have data at only three $\kappa$ values except for
the zero recoil point, it is not practical to test
different functional forms in $m_q$ for the chiral
extrapolation. 
We instead estimate the corresponding error in the form
factors by taking the square of the difference between the
result of the chiral limit and that of the lightest
$\kappa$. 
This gives 10\% for $f_1(v\cdot k_\pi)+f_2(v\cdot k_\pi)$
and 1\% for $f_2(v\cdot k_\pi)$.

The total error is estimated by adding these errors in
quadrature together with the statistical error.
In Figure~\ref{fig:f1f2_physical_with_syst} the form factors
$f_1(v\cdot k_\pi)+f_2(v\cdot k_\pi)$ and 
$f_2(v\cdot k_\pi)$ are plotted with the estimated
systematic uncertainties. 
Numerical results are listed in
Table~\ref{tab:numerical_results}. 


%% file: 7.Comparison.tex
\section{Comparison with other calculations}
\label{sec:Comparison_with_other_calculations}

\subsection{$f_1(v\cdot k_{\pi})$ and $f_2(v\cdot k_{\pi})$}
The Fermilab group calculated the form factors at the $b$
quark mass using the non-relativistic interpretation of the
relativistic lattice action \cite{Fermilab_01}.
A comparison is made with our results for the HQET form
factors 
$f_1(v\cdot k_{\pi})+f_2(v\cdot k_{\pi})$ and
$f_2(v\cdot k_{\pi})$ at the same $\beta$ value employed,
$\beta$=5.9, in Figure~\ref{fig:f1f2_fermilab}.
We find a reasonable agreement for $f_2(v\cdot k_{\pi})$,
but for $f_1(v\cdot k_{\pi})+f_2(v\cdot k_{\pi})$ our data
seem substantially lower.

Since both NRQCD and Fermilab action are the two
variants of the nonrelativistic effective action, 
there should be no fundamental difference in the result.
There can be, however,  two possible reasons for the disagreement.
One is the difference in the renormalization factor.
The other is the difference in various systematic errors 
which arise from the choice of parameters such as 
the lattice size, smearing methods, fitting procedures
and so on.

In order to see he reason of the disagreement, we
plot the form factors at a fixed momentum configuration
$a\boldvec{p}_B=(0,0,0)$ and $a\boldvec{k}_\pi=(1,0,0)$
as a function of the light quark mass in
Figure~\ref{fig:mq_dependence_fermilab}. 
While we find a good agreement for $f_2(v\cdot k_{\pi})$,
our results for $f_1(v\cdot k_{\pi})+f_2(v\cdot k_{\pi})$ is
significantly lower than the Fermilab data.
Furthermore, in the fit of the form
(\ref{eq:global_fit_f1+2}) the chiral limit of our data is
lower than the data at finite $am_q$ as shown in the plot,
in contrast to the Fermilab data, for which the chiral limit
becomes even higher due to a positive curvature.

We note that the renormalization of the vector current
is made using the non-perturbative $Z$ factors of heavy-heavy
and light-light current in the Fermilab analysis
\cite{Fermilab_01}. 
A correction is then made perturbatively for the heavy-light
current. 
Since our results are obtained with an entirely perturbative
matching, systematic errors may enter differently.
The effect of such a `partial' non-perturbative
renormalization for the NRQCD action is an issue of future
investigation. 

We should also note that in
Figure~\ref{fig:mq_dependence_fermilab} the statistical
error in our calculation seems much larger than that in the
Fermilab data, despite much larger statistics in our
calculation.
We suspect that the main reason for the large statistical
error in our data is a larger temporal extent of our
lattice $N_T$=48 compared to $N_T$=32 in the Fermilab work.
The large temporal size and the large distance between
$t_\pi$ and $t_B$ in our simulation renders the extraction of
the ground state contribution very convincing as shown in
Figure~\ref{fig:R}, which seems much better than the
equivalent plot in \cite{Fermilab_01}, but at the same time
the statistical noise exponentially grows as the heavy-light
meson evolves in the temporal direction
\cite{Lepage_lat91,Hashimoto_94}.

\subsection{$f^+(q^2)$ and $f^0(q^2)$}
A comparison of the form factors in the conventional
definition $f^+(q^2)$ and $f^0(q^2)$ is made in
Figure~\ref{fig:fpf0_comparison}.
Results from recent lattice calculations by
APE~\cite{APE_00}, UKQCD~\cite{UKQCD_00} and
Fermilab~\cite{Fermilab_01} are shown in the plot together
with our data.

We find that all data is consistent with each other for
$f^+(q^2)$, while our result is somewhat lower for
$f^0(q^2)$.
Since $f^0(q^2)$ is proportional to 
$f_1(v\cdot k_{\pi})+f_2(v\cdot k_{\pi})$
up to a small correction of $O(v\cdot k_{\pi}/m_B)$, the
disagreement with the Fermilab result is the same one as we
discussed in the previous subsection.

The results of other two groups, APE and UKQCD, are lower 
than the Fermilab result to but still higher than ours.
We note that in their approach an extrapolation in
$1/M_P$ is necessary to predict the $B$ meson form factor
from the simulation results for lighter heavy quarks.
Figure~\ref{fig:phi_comparison} shows such an extrapolation.
The magnitude of
$\Phi_0 (\equiv (\alpha_s(M_P)/\alpha_s(M_B)^{-2/11}f^0\sqrt{M_P})$ 
in their results agrees with ours, but the APE results show 
a negative slope in contrast to a flat $1/M$ dependence of our data, 
leading to the APE value at the physical point considerably higher
than ours. In the relativistic approach, the discretization
error may be magnified toward heavier quarks, since the
discretization error scales as a power of $aM$.
Therefore, the dependence on the heavy quark mass can be
badly distorted.
Furthermore, the heavy quark expansion becomes questionable
for lighter heavy quarks, and the extrapolation with a linear
or quadratic function in $1/M_P$ may not be sufficient.
Such effects are difficult to incorporate in the 
extrapolation, and the systematic error can be
underestimated.


%% file: 8.Phenomenology.tex
\section{Phenomenological Implications}
\label{sec:Phenomenological_Implications}

\subsection{Differential decay rate}
The differential decay rate of the semileptonic
$B^0\rightarrow\pi^- l^+\nu_l$ decay is proportional to the
form factor $f^+(q^2)$ squared, 
\begin{equation}
  \label{eq:differential_decay_rate}
  \frac{1}{|V_{ub}|^2}
  \frac{d\Gamma}{dq^2} =
  \frac{G_F^2}{24\pi^3}
  |\boldvec{k}_{\pi}|^3
  |f^+(q^2)|^2.
\end{equation}
Therefore, if a reliable calculation of the form factor is
available from lattice QCD, the experimental data can be
used to extract the CKM element $|V_{ub}|$.
Our result for the differential decay rate divided by
$|V_{ub}|^2$ is listed in Table~\ref{tab:numerical_results}
and shown in Figure~\ref{fig:diff_decayrate}.

The momentum configuration where data is available is
limited to a large $q^2$ region, 18~GeV$^2$ $<q^2<$
21~GeV$^2$, which corresponds to small recoil momenta.
In the region above 21~GeV$^2$ there is no data point
because of large pion mass in the lattice calculations.
However, the pole dominance near zero recoil region 
(\ref{eq:pole_dominance_f2}) and (\ref{eq:pole_dominance_f+}), 
which is confirmed in part by our lattice calculations, 
should become even better approximation in that region.
Therefore, the theoretical uncertainty is under control in
that large $q^2$ region.
A strategy to determine  $|V_{ub}|$ is, then, to measure the
decay rate in the large $q^2$ region, $q^2 >$ 18~GeV$^2$,
and to use the lattice result
\begin{equation}
  \label{eq:f+_above_18GeV^2}
  \frac{G_F^2}{24\pi^3}
  \int_{18\mbox{\scriptsize GeV}^2}^{q_{max}^2} dq^2
  |\boldvec{k}_{\pi}|^3
  |f^+(q^2)|^2
  = 1.18 \pm 0.37 \pm 0.08 \pm 0.31\;\;
  \mbox{psec}^{-1} 
\end{equation}
The first error is statistical, the second is perturbative
and the last error is the error from discretization and chiral
extrapolation.

\subsection{$D\rightarrow\pi l\nu$ and $D\rightarrow Kl\nu$} 
As we found in
Section~\ref{sec:Heavy_quark_mass_dependence}, 
the $1/M$ correction to the HQET form factors 
$f_1(v\cdot k_{\pi})$ and $f_2(v\cdot k_{\pi})$ is small.
Although our data is only available for large heavy quark
mass $M>$ 3.2~GeV and the charm quark mass is not covered, 
the result suggests that the semileptonic decays of $D$
mesons, $D\rightarrow\pi l\nu$ and $D\rightarrow Kl\nu$, may
be used to constrain the form factors, as proposed by
Burdman \textit{et al.}
\cite{Burdman_Ligeti_Neubert_Nir_94}.

The idea of \cite{Burdman_Ligeti_Neubert_Nir_94} is to
consider a ratio of the differential decay rates of
$B\rightarrow\pi l\nu$ and 
$D\rightarrow\pi l\nu$ at a fixed recoil energy 
$v\cdot k_{\pi}$, then the heavy quark symmetry tells us
that the ratio is unity at the leading order, and the ratio
of CKM elements $|V_{ub}/V_{cd}|$ may be extracted without
model dependence.
The method is, however, not so useful unless the size of
$1/M$ (and higher order) corrections is reliably estimated. 
The lattice calculation could be used to evaluate them, as
we attempt in this work.

In the lattice calculation, the bulk of systematic errors,
especially uncertainty in the perturbative renormalization,
is canceled in the ratio of form factors with different
heavy quark mass.
This idea was extensively used by the Fermilab group
\cite{Fermilab_00,Simone_lat99} in the lattice study of
heavy-to-heavy decay, namely $B\rightarrow D^{(*)}l\nu$, in
which the heavy quark symmetry predicts a stronger
constraint and the form factor is even normalized in the
zero recoil limit up to a correction of $O(1/M^2)$ that can
be calculated on the lattice.
The Fermilab group also considered the ratio for the
heavy-to-light decay \cite{Fermilab_01}.
They calculated the form factors at $b$ and $c$ quark
masses, and found a small but significant mass dependence in
the HQET form factors, which might conflict with our
findings. 
It is, therefore, important to extend our work toward
lighter heavy quarks in order to investigate how the form
factors are modified by the $1/M$ corrections.
We also note that for this purpose the non-relativistic
interpretation of the relativistic lattice action
\cite{El-Khadra_Kronfeld_Mackenzie_97} employed in
\cite{Fermilab_01} is best suited, because lighter heavy
quark can be treated without large systematic errors.


%% file: 9.Conclusion.tex
\section{Conclusions}
\label{sec:Conclusions}

In this paper, we calculated the form factors and the 
differential decay rate for $B\rightarrow\pi l\nu$ on the
quenched lattice using the NRQCD action.
In the HQET form factors $f_1(v\cdot k_{\pi})$
and $f_2(v\cdot k_{\pi})$, the heavy quark mass dependence
appears only in the form of the $1/M$ expansion.
From calculations at several different heavy quark
masses we found that the $1/M$ correction is not significant
for these form factors.
We found that the $B^*$ pole contribution dominates $f^+(q^2)$ 
for small pion recoil energy.
We also showed that the extrapolation to the soft pion limit
suffers from large systematic errors, so that the discrepancies 
between $f^0(q^2_{max})$ and $f_B/f_{\pi}$ in the soft pion relation, 
as seen, in the present simulation is not a serious problem.

In order to avoid the model dependence, we did not assumed any
particular functional form for the form factors.
Instead, we carried out an interpolation in the region where
our data are available.
Although the accessible $q^2$ region is rather limited,
the prediction from chiral effective lagrangian may be used
to extend the prediction toward $q^2_{max}$, and we obtained a 
partially integrated differential decay rate in the region
18~GeV$<q^2<q^2_{max}$.
We obtained 
  $\displaystyle{ \frac{G_F^2}{24\pi^3}
  \int_{18\mbox{\scriptsize GeV}^2}^{q_{max}^2} dq^2
  |\boldvec{k}_{\pi}|^3
  |f^+(q^2)|^2
  = 1.18 \pm 0.37 \pm 0.08 \pm 0.31
  \mbox{psec}^{-1} }$ 
where the first error is statistical error
the second is the error from perturbative calculation 
and the third is the
systematic error from the discretization and chiral extrapolation.

The discretization error
of $O(a^2)$ and the perturbative error are sizable.
The first error can be reduced by performing simulation
at several different lattice spacings and/or using different
lattice actions.
The reduction of the second error is more demanding.
We need a non-perturbative
renormalization to remove it.
Another important source of uncertainty, which we did not include, is 
the quenched approximation, whose effect can be estimated only with
simulations including dynamical quarks.
We are planning future studies in these directions.


%% file: Reference.tex
%
%


%% file: Table.tex
\begin{table}
  \begin{center}
    \begin{tabular}{ccccc} 
      $aM_0$ & $n$ & 
      $\rho_{V_4}^{(0)}$ & 
      $\rho_{V_4}^{(1)}$ & 
      $\rho_{V_4}^{(2)}$ \\
      \hline         
      10.0 & 2 & $-$0.562 & $-$0.572 & $-$0.421\\
       5.0 & 2 & $-$0.554 & $-$0.571 &  0.205\\
       3.0 & 2 & $-$0.540 & $-$0.582 &  0.446\\
       2.1 & 3 & $-$0.529 & $-$0.604 &  0.559\\
       1.3 & 3 & $-$0.509 & $-$0.629 &  0.657\\
    \end{tabular}
    \caption{Renormalization constants for $V_4$.}
    \label{tab:V4}
  \end{center}
\end{table}

\begin{table}
  \begin{center}
    \begin{tabular}{ccccccc} 
      $aM_0$ & $n$ & 
      $\rho_{V_k}^{(0)}$ & 
      $\rho_{V_k}^{(1)}$ & 
      $\rho_{V_k}^{(2)}$ &
      $\rho_{V_k}^{(3)}$ &
      $\rho_{V_k}^{(4)}$ \\
      \hline         
      10.0 & 2 & $-$1.250 & 0.366   & 13.705 & 0.983 & 1.047\\
       5.0 & 2 & $-$1.087 & 0.232   &  4.678 & 0.881 & 0.977\\
       3.0 & 2 & $-$0.915 & 0.091   &  1.605 & 0.774 & 0.893\\
       2.1 & 3 & $-$0.772 &$-$0.049 &  0.594 & 0.690 & 0.812\\
       1.3 & 3 & $-$0.546 &$-$0.235 &  0.188 & 0.587 & 0.668\\
    \end{tabular}
    \caption{Renormalization constants for $V_k$.}
    \label{tab:Vk}
  \end{center}
\end{table}

\begin{table}
  \begin{center}
    \begin{tabular}{ccc} 
      $aM_0$ & $a_h$ & $b_h$ \\
      \hline
      10.0   & 0.16  & 1.50 \\
      5.0    & 0.28  & 1.12 \\
      3.0    & 0.29  & 1.07 \\
      2.1    & 0.30  & 1.06 \\
      1.3    & 0.31  & 1.06 \\
    \end{tabular}
    \caption{Smearing parameters for the heavy-light meson.}
    \label{tab:smear_parameters}
  \end{center}
\end{table}

\begin{table}
  \begin{tabular}{cccc}
    id. &
    $\boldvec{p}_B$ & 
    $\boldvec{k}_{\pi}$ &
    $\boldvec{q}$ \\
    \hline
    p000.q000 & (0,0,0) & (0,0,0)    & (0,0,0)\\
    p100.q100 & (1,0,0) & (0,0,0)    & (1,0,0)\\
    p100.q000 & (1,0,0) & (1,0,0)    & (0,0,0)\\
    p100.q110 & (1,0,0) & (0,$-1$,0) & (1,1,0)\\
    p000.q100 & (0,0,0) & ($-1$,0,0) & (1,0,0)\\
    p100.q200 & (1,0,0) & ($-1$,0,0) & (2,0,0)  
  \end{tabular}
  \caption{List of momenta for which the three-point
    correlator is measured.
    Three-momentum is given in the unit of $2\pi/La$.
    'id.' will be used in the tables of numerical results.
    }
  \label{tab:momenta}
\end{table}

\begin{table}
  \begin{tabular}{lllllllll}
    \multicolumn{9}{l}{$aM_0$=5.0}\\
    \hline
    id. & 
    $\hat{V}_4^{(0)}$ &  $\hat{V}_4^{(1)}$  &
    $\hat{V}_4^{(2)}$ &
    $\hat{V}_1^{(0)}$ &  $\hat{V}_1^{(1)}$ &
    $\hat{V}_1^{(2)}$ & $\hat{V}_1^{(3)}$ &
    $\hat{V}_1^{(4)}$\\
    \hline
    p000.q000& 2.059(44)& 0.1874(53)&-0.1874(53)& & & & & \\
    p100.q100& 2.052(52)& 0.2304(72)&-0.1862(60)
    & 0.010(10)&-0.660(17)& 0.001(3)&-0.713(18)&-0.053(4)\\
    p100.q000& 1.79(17)&-0.023(28)& 0.023(28)
    & 0.876(90)&-0.496(53)& 0.013(20)&-0.241(26)& 0.241(26)\\
    p100.q110& 1.645(64)& 0.272(15)& 0.0602(86)
    &-0.740(30)&-0.121(26)& 0.004(5)&-0.358(20)&-0.266(10)\\
    p000.q100& 1.554(56)& 0.216(13)& 0.0603(97)
    &-0.751(29)&-0.063(23)& 0.011(7)&-0.321(19)&-0.268(11)\\
    p100.q200& 1.34(11)& 0.487(43)& 0.073(20)
    &-0.697(62)&-0.547(70)&0.042(16)&-0.765(71)&-0.263(25)\\
    \hline\hline
    \multicolumn{9}{l}{$aM_0$=3.0}\\
    \hline
    id. & 
    $\hat{V}_4^{(0)}$ &  $\hat{V}_4^{(1)}$  &
    $\hat{V}_4^{(2)}$ &
    $\hat{V}_1^{(0)}$ &  $\hat{V}_1^{(1)}$ &
    $\hat{V}_1^{(2)}$ & $\hat{V}_1^{(3)}$ &
    $\hat{V}_1^{(4)}$\\
    \hline
    p000.q000& 2.087(33)& 0.217(41)&-0.2173(41)&&&&&\\
    p100.q100& 2.066(39)& 0.2802(60)&-0.2126(47)
    &0.031(8)&-0.632(12)& 0.016(2)&-0.697(13)&-0.081(3)\\
    p100.q000& 1.81(16)& 0.008(23)&-0.008(23)
    &0.877(80)&-0.416(44)& 0.003(16)&-0.207(22)& 0.207(22)\\
    p100.q110& 1.621(57)& 0.302(13)& 0.0426(66)
    &-0.704(26)&-0.152(22)& 0.034(4)&-0.351(17)&-0.274(9)\\
    p000.q100& 1.547(50)& 0.231(11)& 0.0390(73)
    &-0.710(25)&-0.096(18)& 0.0393(53)&-0.325(16)&-0.268(9)\\
    p100.q200& 1.343(94)& 0.495(34)& 0.073(15)
    &-0.627(46)&-0.544(53)& 0.076(13)&-0.762(57)&-0.296(20)\\
    \hline\hline
    \multicolumn{9}{l}{$aM_0$=2.1}\\
    \hline
    id. & 
    $\hat{V}_4^{(0)}$ &  $\hat{V}_4^{(1)}$  &
    $\hat{V}_4^{(2)}$ &
    $\hat{V}_1^{(0)}$ &  $\hat{V}_1^{(1)}$ &
    $\hat{V}_1^{(2)}$ & $\hat{V}_1^{(3)}$ &
    $\hat{V}_1^{(4)}$\\
    \hline
    p000.q000& 2.108(30)& 0.2552(41)&-0.2552(41)&&&&&\\
    p100.q100& 2.072(35)& 0.3354(61)&-0.2469(46)
    &0.057(7)&-0.606(10)& 0.030(2)&-0.680(11)&-0.105(3)\\
    p100.q000& 1.85(16)& 0.050(22)&-0.050(22)
    &0.898(73)&-0.347(39)&-0.008(14)&-0.178(19)& 0.178(19)\\
    p100.q110& 1.607(51)& 0.341(12)& 0.0172(58)
    &-0.683(23)&-0.184(18)& 0.060(4)&-0.350(15)&-0.278(8)\\
    p000.q100& 1.555(47)& 0.254(10)& 0.0129(65)
    &-0.686(22)&-0.130(15)& 0.064(5)&-0.333(14)&-0.268(8)\\
    p100.q200& 1.321(86)& 0.504(32)& 0.067(15)
    &-0.556(41)&-0.551(48)& 0.117(16)&-0.749(51)&-0.315(19)\\
    \hline\hline
    \multicolumn{9}{l}{$aM_0$=1.3}\\
    \hline
    id. & 
    $\hat{V}_4^{(0)}$ &  $\hat{V}_4^{(1)}$  &
    $\hat{V}_4^{(2)}$ &
    $\hat{V}_1^{(0)}$ &  $\hat{V}_1^{(1)}$ &
    $\hat{V}_1^{(2)}$ & $\hat{V}_1^{(3)}$ &
    $\hat{V}_1^{(4)}$\\
    \hline
    p000.q000& 2.140(25)& 0.3278(40)&-0.3278(40)&&&&&\\
    p100.q100& 2.066(34)& 0.4323(74)&-0.3087(54)
    &0.103(6)&-0.556(10)& 0.057(2)&-0.644(10)&-0.146(3)\\
    p100.q000& 1.94(16)& 0.136(24)&-0.136(24)
    &0.943(66)&-0.229(34)&-0.030(12)&-0.130(16)& 0.130(16)\\
    p100.q110& 1.519(45)& 0.399(14)&-0.0227(63)
    &-0.636(23)&-0.234(16)& 0.103(5)&-0.338(13)&-0.269(8)\\
    p000.q100& 1.581(45)& 0.304(10)&-0.0361(63)
    &-0.655(19)&-0.189(14)& 0.110(5)&-0.348(13)&-0.269(7)\\
    p100.q200& 1.278(71)& 0.551(30)& 0.049(11)
    &-0.478(31)&-0.564(39)& 0.180(14)&-0.733(41)&-0.349(17)
  \end{tabular}
  \caption{
    Matrix elements $\hat{V}_\mu^{(i)}$ at $\kappa$=0.13630.
    The first column represents the momentum configuration
    as defined in Table~\ref{tab:momenta}.
  \label{tab:V.k1}
    }
\end{table}

\begin{table}
  \begin{tabular}{lllllllll}
    \multicolumn{9}{l}{$aM_0$=5.0}\\
    \hline
    id. & 
    $\hat{V}_4^{(0)}$ &  $\hat{V}_4^{(1)}$  &
    $\hat{V}_4^{(2)}$ &
    $\hat{V}_1^{(0)}$ &  $\hat{V}_1^{(1)}$ &
    $\hat{V}_1^{(2)}$ & $\hat{V}_1^{(3)}$ &
    $\hat{V}_1^{(4)}$\\
    \hline
    p000.q000& 2.222(61)& 0.2754(88)&-0.2754(88)&&&&&\\
    p100.q100& 2.211(70)& 0.326(11)&-0.2756(99)
    &-0.005(13)&-0.708(23)&-0.009(4)&-0.772(24)&-0.056(5)\\
    p100.q000& 1.76(30)& 0.051(40)&-0.051(40)
    &1.11(15)&-0.523(82)& 0.071(28)&-0.224(41)& 0.224(41)\\
    p100.q110& 1.71(11)& 0.338(31)& 0.036(15)
    &-0.843(59)&-0.126(46)&-0.056(10)&-0.408(36)&-0.243(19)\\
    p000.q100& 1.50(10)& 0.262(26)& 0.038(17)
    &-0.826(56)&-0.037(41)&-0.037(13)&-0.322(34)&-0.249(20)\\
    p100.q200& 1.31(19)& 0.556(78)& 0.074(30)
    &-0.81(11)&-0.53(12)&-0.008(26)&-0.77(12)&-0.229(40)\\
    \hline\hline
    \multicolumn{9}{l}{$aM_0$=3.0}\\
    \hline
    id. & 
    $\hat{V}_4^{(0)}$ &  $\hat{V}_4^{(1)}$  &
    $\hat{V}_4^{(2)}$ &
    $\hat{V}_1^{(0)}$ &  $\hat{V}_1^{(1)}$ &
    $\hat{V}_1^{(2)}$ & $\hat{V}_1^{(3)}$ &
    $\hat{V}_1^{(4)}$\\
    \hline
    p000.q000& 2.242(48)& 0.3050(72)&-0.3050(72)&&&&&\\
    p100.q100& 2.217(54)& 0.375(10)&-0.3003(80)
    &0.019(11)&-0.679(17)& 0.009(3)&-0.755(18)&-0.085(4)\\
    p100.q000& 1.81(27)& 0.063(31)&-0.063(31)
    &1.06(12)&-0.408(65)& 0.058(19)&-0.175(32)& 0.175(32)\\
    p100.q110& 1.667(97)& 0.362(27)& 0.020(12)
    &-0.789(50)&-0.151(38)&-0.020(7)&-0.390(30)&-0.256(15)\\
    p000.q100& 1.485(89)& 0.268(21)& 0.020(13)
    &-0.771(47)&-0.076(32)&-0.004(10)&-0.326(29)&-0.247(15)\\
    p100.q200& 1.32(15)& 0.564(59)& 0.062(21)
    &-0.726(81)&-0.519(87)& 0.034(18)&-0.770(93)&-0.285(30)\\
    \hline\hline
    \multicolumn{9}{l}{$aM_0$=2.1}\\
    \hline
    id. & 
    $\hat{V}_4^{(0)}$ &  $\hat{V}_4^{(1)}$  &
    $\hat{V}_4^{(2)}$ &
    $\hat{V}_1^{(0)}$ &  $\hat{V}_1^{(1)}$ &
    $\hat{V}_1^{(2)}$ & $\hat{V}_1^{(3)}$ &
    $\hat{V}_1^{(4)}$\\
    \hline
    p000.q000& 2.254(42)& 0.3434(67)&-0.3434(67)&&&&&\\
    p100.q100& 2.214(50)& 0.431(10)&-0.3345(81)
    &0.050(10)&-0.650(15)& 0.024(2)&-0.737(16)&-0.111(4)\\
    p100.q000& 1.85(25)& 0.092(30)&-0.092(30)
    &1.05(11)&-0.330(56)& 0.046(16)&-0.141(27)& 0.141(27)\\
    p100.q110& 1.634(86)& 0.399(25)&-0.005(10)
    &-0.765(43)&-0.183(32)& 0.013(6)&-0.382(27)&-0.263(13)\\
    p000.q100& 1.495(83)& 0.287(20)&-0.003(11)
    &-0.743(42)&-0.115(28)& 0.024(8)&-0.337(26)&-0.247(13)\\
    p100.q200& 1.31(12)& 0.590(51)& 0.055(17)
    &-0.679(61)&-0.539(71)& 0.076(15)&-0.777(78)&-0.314(26)\\
    \hline\hline
    \multicolumn{9}{l}{$aM_0$=1.3}\\
    \hline
    id. & 
    $\hat{V}_4^{(0)}$ &  $\hat{V}_4^{(1)}$  &
    $\hat{V}_4^{(2)}$ &
    $\hat{V}_1^{(0)}$ &  $\hat{V}_1^{(1)}$ &
    $\hat{V}_1^{(2)}$ & $\hat{V}_1^{(3)}$ &
    $\hat{V}_1^{(4)}$\\
    \hline
    p000.q000& 2.276(35)& 0.4170(66)&-0.4170(66)&&&&&\\
    p100.q100& 2.193(43)& 0.527(10)&-0.3956(80)
    &0.1038(80)&-0.595(12)& 0.0531(20)&-0.697(13)&-0.1557(37)\\
    p100.q000& 1.91(23)& 0.146(35)&-0.146(35)
    &1.02(10)&-0.215(47)& 0.019(17)&-0.098(23)& 0.098(23)\\
    p100.q110& 1.486(72)& 0.443(25)&-0.041(11)
    &-0.701(40)&-0.225(27)& 0.0644(74)&-0.352(23)&-0.254(11)\\
    p000.q100& 1.529(76)& 0.332(19)&-0.049(11)
    &-0.711(34)&-0.182(23)& 0.0754(82)&-0.356(22)&-0.250(12)\\
    p100.q200& 1.28(10)& 0.620(47)& 0.037(14)
    &-0.575(45)&-0.559(56)& 0.149(15)&-0.765(62)&-0.355(23)
  \end{tabular}
  \caption{
    Matrix elements $\hat{V}_\mu^{(i)}$ at $\kappa$=0.13711.
  \label{tab:V.k2}
    }
\end{table}

\begin{table}
  \begin{tabular}{lllllllll}
    \multicolumn{9}{l}{$aM_0$=5.0}\\
    \hline
    id. & 
    $\hat{V}_4^{(0)}$ &  $\hat{V}_4^{(1)}$  &
    $\hat{V}_4^{(2)}$ &
    $\hat{V}_1^{(0)}$ &  $\hat{V}_1^{(1)}$ &
    $\hat{V}_1^{(2)}$ & $\hat{V}_1^{(3)}$ &
    $\hat{V}_1^{(4)}$\\
    \hline
    p000.q000& 2.433(92)& 0.382(15)&-0.382(15)&&&&&\\
    p100.q100& 2.42(10)& 0.448(21)&-0.385(17)
    &-0.031(21)&-0.771(34)&-0.020(60)&-0.850(37)&-0.059(8)\\
    p100.q000& 1.75(55)& 0.118(75)&-0.118(75)
    &1.47(31)&-0.57(15)& 0.170(62)&-0.197(77)& 0.197(77)\\
    p100.q110& 1.74(19)& 0.390(63)& 0.019(29)
    &-0.94(11)&-0.130(85)&-0.120(23)&-0.461(65)&-0.203(35)\\
    p000.q100& 1.33(17)& 0.292(50)& 0.014(32)
    &-0.85(11)& 0.007(75)&-0.081(28)&-0.288(63)&-0.224(36)\\
    p100.q200& 1.36(33)& 0.70(16)& 0.089(53)
    &-1.06(24)&-0.55(22)&-0.061(53)&-0.83(21)&-0.216(74)\\
    \hline\hline
    \multicolumn{9}{l}{$aM_0$=3.0}\\
    \hline
    id. & 
    $\hat{V}_4^{(0)}$ &  $\hat{V}_4^{(1)}$  &
    $\hat{V}_4^{(2)}$ &
    $\hat{V}_1^{(0)}$ &  $\hat{V}_1^{(1)}$ &
    $\hat{V}_1^{(2)}$ & $\hat{V}_1^{(3)}$ &
    $\hat{V}_1^{(4)}$\\
    \hline
    p000.q000& 2.439(71)& 0.411(12)&-0.411(12)&&&&&\\
    p100.q100& 2.413(78)& 0.494(17)&-0.406(13)
    &-0.001(17)&-0.738(24)&-0.000(4)&-0.830(27)&-0.091(6)\\
    p100.q000& 1.76(50)& 0.111(60)&-0.111(60)
    &1.38(25)&-0.40(11)& 0.145(45)&-0.127(60)& 0.127(60)\\
    p100.q110& 1.69(16)& 0.407(53)& 0.003(22)
    &-0.864(96)&-0.142(68)&-0.076(16)&-0.430(54)&-0.230(28)\\
    p000.q100& 1.31(15)& 0.283(42)& 0.004(24)
    &-0.785(91)&-0.043(58)&-0.038(19)&-0.294(53)&-0.218(27)\\
    p100.q200& 1.37(26)& 0.70(12)& 0.057(38)
    &-0.92(16)&-0.55(15)&-0.006(34)&-0.83(16)&-0.284(56)\\
    \hline\hline
    \multicolumn{9}{l}{$aM_0$=2.1}\\
    \hline
    id. & 
    $\hat{V}_4^{(0)}$ &  $\hat{V}_4^{(1)}$  &
    $\hat{V}_4^{(2)}$ &
    $\hat{V}_1^{(0)}$ &  $\hat{V}_1^{(1)}$ &
    $\hat{V}_1^{(2)}$ & $\hat{V}_1^{(3)}$ &
    $\hat{V}_1^{(4)}$\\
    \hline
    p000.q000& 2.438(61)& 0.449(11)&-0.449(11)&&&&&\\
    p100.q100& 2.392(72)& 0.548(17)&-0.439(13)
    &0.037(14)&-0.705(22)& 0.017(3)&-0.807(24)&-0.119(5)\\
    p100.q000& 1.81(46)& 0.132(59)&-0.132(59)
    &1.33(21)&-0.305(97)& 0.125(37)&-0.086(51)& 0.086(51)\\
    p100.q110& 1.63(14)& 0.441(49)&-0.023(19)
    &-0.837(83)&-0.169(57)&-0.036(12)&-0.410(47)&-0.244(24)\\
    p000.q100& 1.33(14)& 0.299(39)&-0.014(22)
    &-0.760(79)&-0.088(49)&-0.009(15)&-0.310(47)&-0.217(24)\\
    p100.q200& 1.35(22)& 0.72(10)& 0.047(31)
    &-0.85(12)&-0.572(12)& 0.043(27)&-0.85(13)&-0.315(47)\\
    \hline\hline
    \multicolumn{9}{l}{$aM_0$=1.3}\\
    \hline
    id. & 
    $\hat{V}_4^{(0)}$ &  $\hat{V}_4^{(1)}$  &
    $\hat{V}_4^{(2)}$ &
    $\hat{V}_1^{(0)}$ &  $\hat{V}_1^{(1)}$ &
    $\hat{V}_1^{(2)}$ & $\hat{V}_1^{(3)}$ &
    $\hat{V}_1^{(4)}$\\
    \hline
    p000.q000& 2.445(51)& 0.525(11)&-0.525(11)&&&&&\\
    p100.q100& 2.350(61)& 0.643(17)&-0.499(12)
    &0.102(12)&-0.643(17)& 0.050(3)&-0.762(20)&-0.169(5)\\
    p100.q000& 1.86(41)& 0.163(68)&-0.163(68)
    &1.21(19)&-0.190(83)& 0.088(35)&-0.049(44)& 0.049(44)\\
    p100.q110& 1.43(11)& 0.476(47)&-0.058(21)
    &-0.751(72)&-0.200(46)& 0.024(12)&-0.355(40)&-0.242(20)\\
    p000.q100& 1.38(13)& 0.341(37)&-0.054(22)
    &-0.742(63)&-0.160(41)& 0.042(14)&-0.337(40)&-0.222(21)\\
    p100.q200& 1.35(17)& 0.741(86)& 0.023(26)
    &-0.711(87)&-0.610(96)& 0.126(25)&-0.85(10)&-0.367(42)
  \end{tabular}
  \caption{
    Matrix elements $\hat{V}_\mu^{(i)}$ at $\kappa$=0.13769.
  \label{tab:V.k3}
    }
\end{table}

\begin{table}
  \begin{tabular}{llll}
    $aM_0$ & 
    $\hat{V}_4^{(0)}$ & $\hat{V}_4^{(1)}$ & $\hat{V}_4^{(2)}$ \\ 
    \hline
    5.0 & 2.78(15) & 0.551(31)&-0.551(31)\\
    3.0 & 2.76(11) & 0.578(23)&-0.578(23)\\
    2.1 & 2.74(10) & 0.615(22)&-0.615(22)\\
    1.3 & 2.721(85)& 0.690(20)&-0.690(20)
  \end{tabular}
  \caption{
    Matrix element $\hat{V}_4^{(i)}$ at
    $\kappa$=0.13816 for the zero recoil configuration
    (p000.q000).
    }
  \label{tab:V.k4}
\end{table}

\begin{table}
  \begin{tabular}{lllll}
    $\kappa$ & 0.13630 & 0.13711 & 0.13769 & 0.13816\\
    \hline
    (0,0,0) & 0.48816(62) & 0.40756(68) & 0.34005(78) & 0.2723(11)\\
    (1,0,0) & 0.6209(28)  & 0.5578(44)  & 0.5065(73)  & 0.459(15) \\
    (1,1,0) & 0.7424(43)  & 0.6967(71)  & 0.654(15)   & 0.626(32)
  \end{tabular}
  \caption{
    Pion energy $aE_{\pi}(\boldvec{k}_{\pi})$ for
    spatial momenta (0,0,0), (1,0,0), (1,1,0) in unit of
    $2\pi/La$.} 
  \label{tab:pion_energy}
\end{table}

\begin{table}
  \begin{tabular}{lllll}
    $\kappa$ & 0.13630 & 0.13711 & 0.13769 & 0.13816 \\
    \hline
    $aM_0$=5.0 & 5.5702(21) & 5.5476(29) & 5.5318(39) & 5.5203(58)\\
    $aM_0$=3.0 & 3.6606(15) & 3.6376(20) & 3.6212(27) & 3.6088(40)\\
    $aM_0$=2.1 & 2.7992(12) & 2.7758(16) & 2.7588(22) & 2.7455(32)\\
    $aM_0$=1.3 & 2.0301(10) & 2.0058(13) & 1.9883(17) & 1.9741(25)
  \end{tabular}
  \caption{
    Heavy-light meson mass $aM_B$ evaluated using
    (\ref{eq:heavy-light_meson_mass}) with $\alpha_V(1/a)$.
    }
  \label{tab:M_B}
\end{table}

\begin{table}
  \begin{tabular}{lllll}
    \multicolumn{5}{l}{$aM_0$=5.0}\\
    \hline
    $\kappa$ & 0.13630 & 0.13711 & 0.13769 & 0.13816\\
    (0,0,0) & 0.5887(21) & 0.5661(29) & 0.5503(39) & 0.5388(58)\\
    (1,0,0) & 0.6062(15) & 0.5821(22) & 0.5668(29) & 0.5554(40)\\
    \hline\hline
    \multicolumn{5}{l}{$aM_0$=3.0}\\
    \hline
    $\kappa$ & 0.13630 & 0.13711 & 0.13769 & 0.13816\\
    (0,0,0) & 0.5830(15) & 0.5600(20) & 0.5436(27) & 0.5312(40)\\
    (1,0,0) & 0.6071(13) & 0.5826(17) & 0.5665(23) & 0.5546(32)\\
    \hline\hline
    \multicolumn{5}{l}{$aM_0$=2.1}\\
    \hline
    $\kappa$ & 0.13630 & 0.13711 & 0.13769 & 0.13816\\
    (0,0,0) & 0.5750(12) & 0.5515(16) & 0.5346(22) & 0.5212(32)\\
    (1,0,0) & 0.6054(12) & 0.5804(15) & 0.5638(20) & 0.5512(28)\\
    \hline\hline
    \multicolumn{5}{l}{$aM_0$=1.3}\\
    \hline
    $\kappa$ & 0.13630 & 0.13711 & 0.13769 & 0.13816\\
    (0,0,0) & 0.5571(10) & 0.5328(13) & 0.5152(17) & 0.5010(25)\\
    (1,0,0) & 0.5984(12) & 0.5711(18) & 0.5532(24) & 0.5387(35)
  \end{tabular}
  \caption{Binding energy of the heavy-light meson
    $aE_{\mathrm{bin}}(\boldvec{p}_B)$ for spatial momenta
    (0,0,0) and (1,0,0) in unit of $2\pi/La$.}
  \label{tab:B_binding_energy}
\end{table}

\begin{table}
  \begin{tabular}{cccccc}
    $C_{1+2}^{(000)}$ & $C_{1+2}^{(010)}$ &
    $C_{1+2}^{(100)}$ & $C_{1+2}^{(001)}$ & 
    $C_{1+2}^{(011)}$ & $C_{1+2}^{(002)}$ \\
    \hline
    0.413(74) & 3.79(97) & $-$0.019(31) & 
    $-$0.53(66) & $-$3.3(13.8) & 0.7(2.3) \\
    0.392(75) & 3.94(99) & 0.070(29) &
    $-$0.59(66) & $-$3.5(13.8) & 0.7(2.3) \\
    \hline\hline
    $C_2^{(000)}$ & $C_2^{(010)}$ & $C_2^{(100)}$ &
    $C_2^{(001)}$ & & \\
    \hline
    0.311(47) & 1.06(1.11) &    0.035(37) & $-$0.06(40) & &\\
    0.347(50) & 0.99(1.14) & $-$0.020(37) & $-$0.04(40) & &
  \end{tabular}
  \caption{
    Global fit parameters in the form 
    (\ref{eq:global_fit_f1+2})--(\ref{eq:global_fit_f2}).
    In each column, top and bottom numbers correspond to the
    result with $\alpha_V(1/a)$ and $\alpha_V(\pi/a)$.
    }
  \label{tab:global_fit_parameters}
\end{table}

\begin{table}
  \begin{center}
    \begin{tabular}{cccccccc}
      $v\cdot k_{\pi}$ & $q^2$ &
      $f_1(v\cdot k_\pi)+f_2(v\cdot k_\pi)$ &
      $f_2(v\cdot k_\pi)$ &
      $f^0(q^2)$ & $f^+(q^2)$ & 
      $1/|V_{ub}|^2d\Gamma/dq^2$ &\\
      (GeV) & (GeV$^2$) & (GeV$^{1/2}$) & (GeV$^{1/2}$) &  & 
      & (psec$^{-1}$GeV$^{-2}$) & \\
      \hline
      0.1435 & 26.37 & 0.98(23) & 0.45(20)  & 0.84(18)  & 7.4(3.1) & 0.0017(15) & \\  
      0.1913 & 25.87 & 0.95(20) & 0.44(19)  & 0.80(16)  & 5.5(2.2) & 0.021(17) & \\
      0.2392 & 25.36 & 0.91(18) & 0.44(18)  & 0.76(15)  & 4.4(1.6) & 0.042(31) & \\
      0.2870 & 24.86 & 0.87(17) & 0.44(17)  & 0.73(14)  & 3.7(1.2) & 0.062(41) & \\
      0.3348 & 24.35 & 0.83(17) & 0.44(15)  & 0.70(14)  & 3.18(98) & 0.081(50) & \\
      0.3827 & 23.85 & 0.79(17) & 0.44(14)  & 0.66(14)  & 2.78(78) & 0.099(56) & \\
      0.4305 & 23.34 & 0.76(18) & 0.43(13)  & 0.63(14)  & 2.46(64) & 0.115(60) & \\
      0.4783 & 22.84 & 0.73(18) & 0.43(12)  & 0.61(15)  & 2.21(53) & 0.131(62) & \\
      0.5262 & 22.33 & 0.69(18) & 0.43(11)  & 0.58(15)  & 2.00(44) & 0.146(64) & \\
      0.5740 & 21.83 & 0.66(18) & 0.43(10)  & 0.55(14)  & 1.82(37) & 0.161(65) & \\
      0.6218 & 21.32 & 0.64(18) & 0.428(93) & 0.53(14)  & 1.67(31) & 0.174(65) & \\
      0.6697 & 20.82 & 0.61(17) & 0.426(88) & 0.51(14)  & 1.54(27) & 0.187(66) & $\ast$ \\
      0.7175 & 20.31 & 0.58(17) & 0.424(85) & 0.49(13)  & 1.43(24) & 0.199(68) & $\ast$ \\
      0.7653 & 19.81 & 0.56(15) & 0.422(84) & 0.47(12)  & 1.33(22) & 0.210(71) & $\ast$ \\
      0.8132 & 19.30 & 0.54(14) & 0.421(86) & 0.45(11)  & 1.24(21) & 0.221(76) & $\ast$ \\
      0.8610 & 18.80 & 0.52(13) & 0.419(90) & 0.435(98) & 1.16(21) & 0.231(84) & $\ast$ \\
      0.9088 & 18.29 & 0.50(12) & 0.417(97) & 0.421(92) & 1.09(21) & 0.240(94) & $\ast$ \\
      0.9567 & 17.79 & 0.48(11) & 0.42(10)  & 0.407(92) & 1.03(22) & 0.25(11) & $\ast$ \\
      1.0045 & 17.28 & 0.47(13) & 0.41(11)  & 0.40(10)  & 0.97(23) & 0.26(13) & \\
      1.0523 & 16.78 & 0.45(15) & 0.41(12)  & 0.38(12)  & 0.92(25) & 0.27(15) & \\
      1.1002 & 16.27 & 0.44(19) & 0.41(14)  & 0.38(15)  & 0.87(27) & 0.27(17) & \\
      1.1480 & 15.77 & 0.43(24) & 0.41(15)  & 0.37(18)  & 0.83(29) & 0.28(20) & \\
    \end{tabular}
  \end{center}
  \caption{
    Numerical results for the form factors and for the
    differential decay rate.
    Error contains the statistical and estimated systematic
    uncertainties. 
    Results with a $\ast$ symbol in the last column are
    those obtained by interpolating the lattice data in
    $v\cdot k_{\pi}$, while others involve an extrapolation.
    }
  \label{tab:numerical_results}
\end{table}


%% file: Figure.tex
\begin{figure}
  \begin{center}
    \begin{tabular}{ll}
      \psfig{file=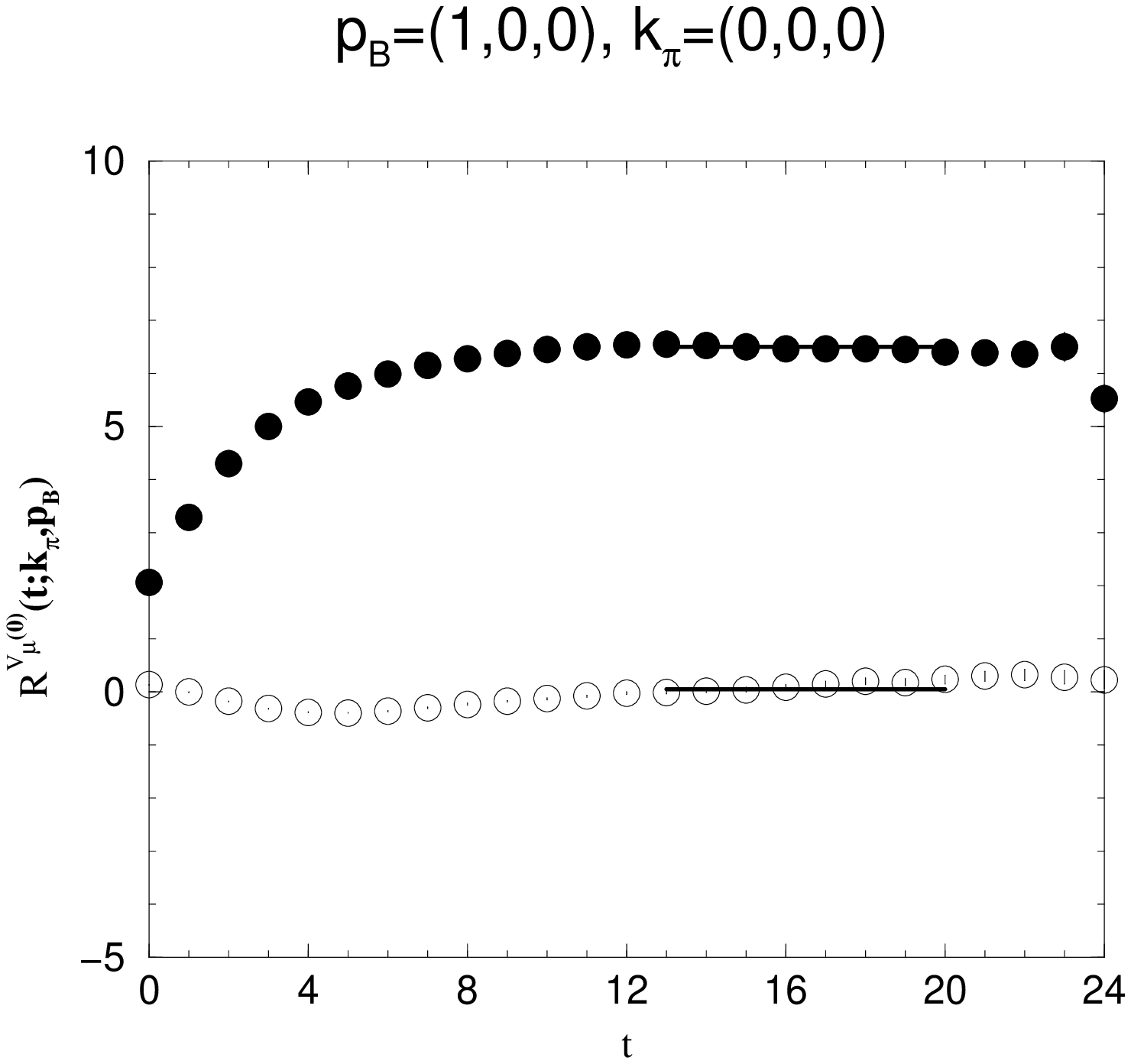,width=7cm,clip=,silent=} &
      \psfig{file=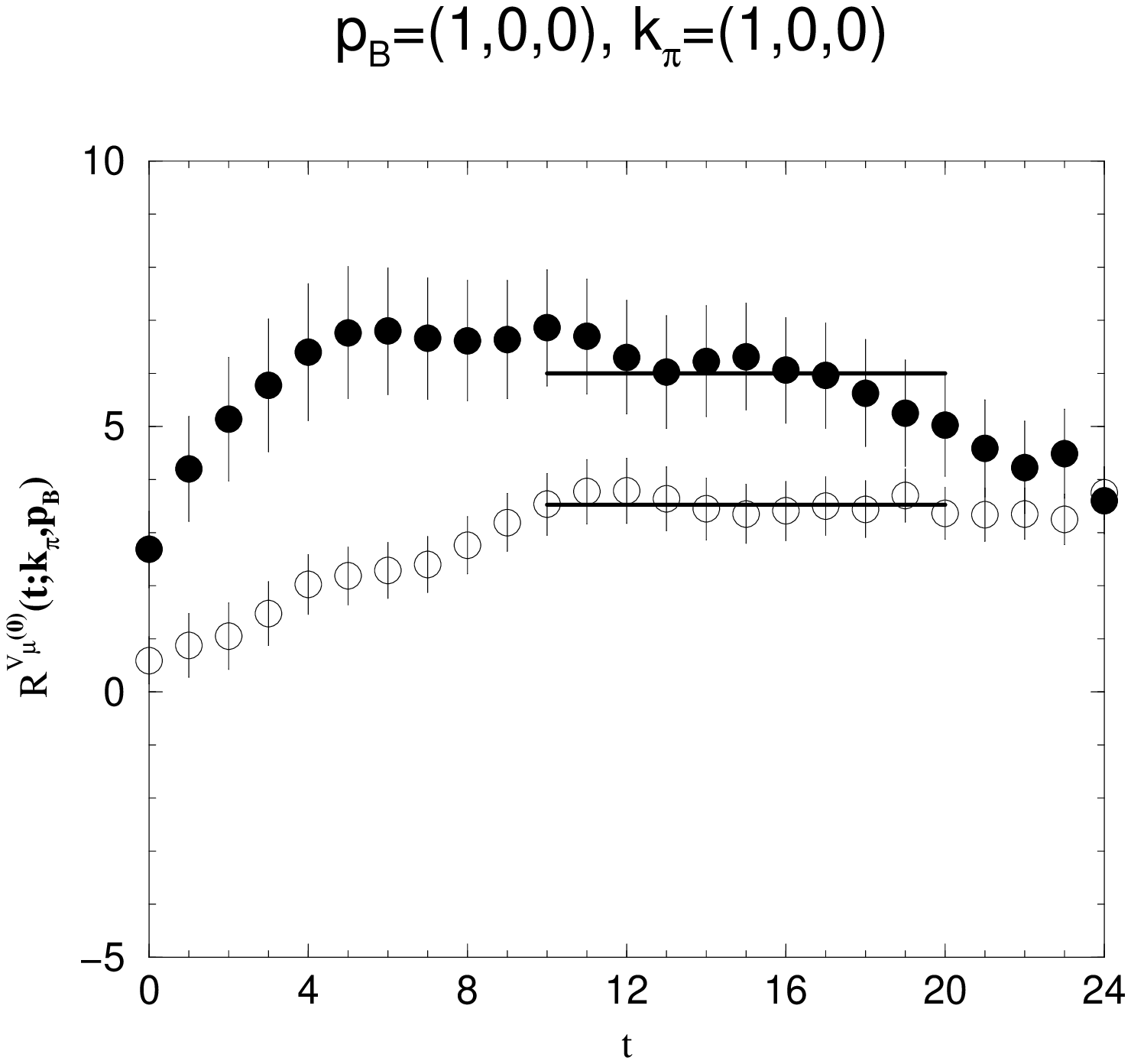,width=7cm,clip=,silent=} \\
      \psfig{file=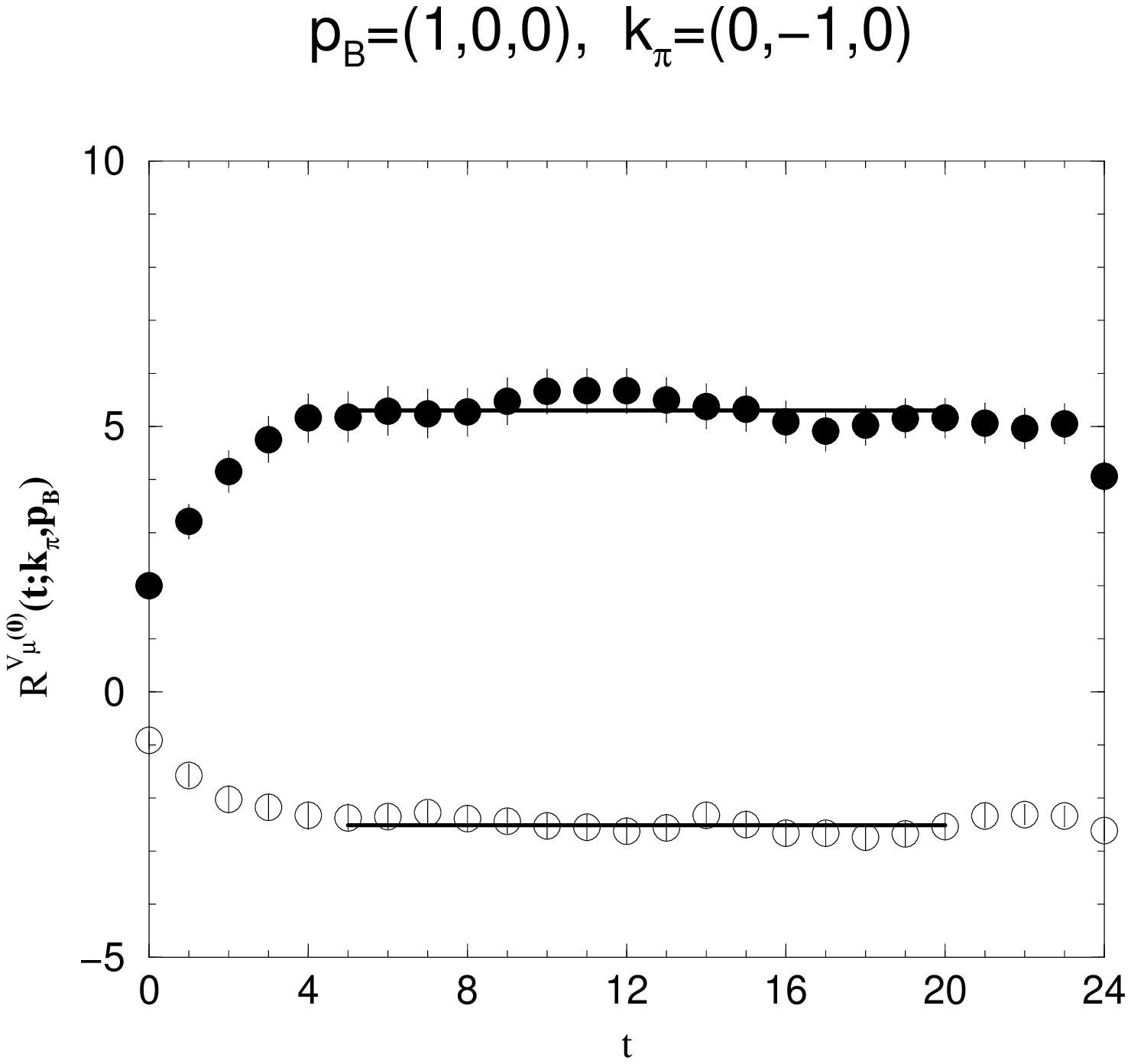,width=7cm,clip=,silent=} &
      \psfig{file=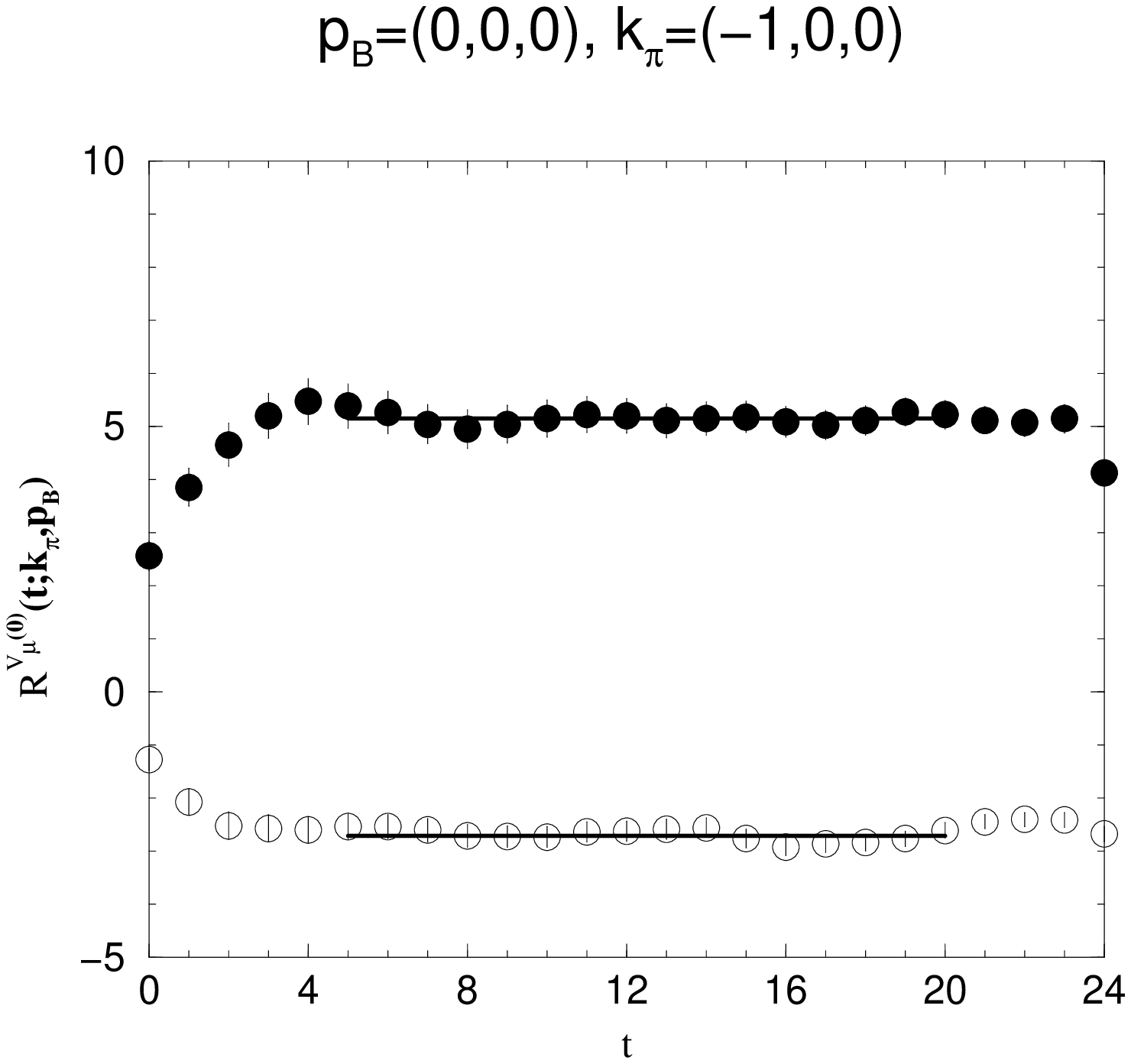,width=7cm,clip=,silent=} \\
      \psfig{file=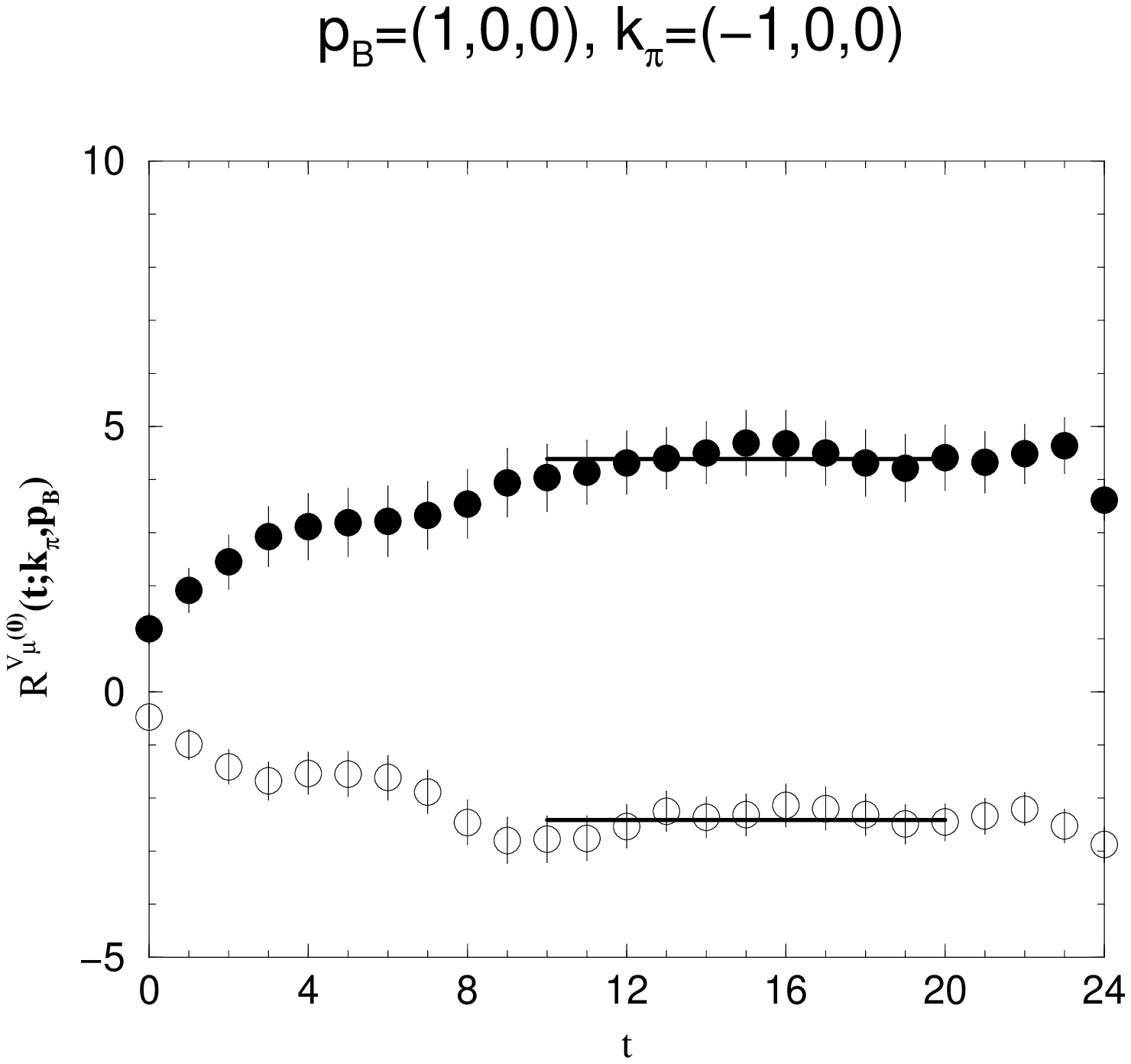,width=7cm,clip=,silent=} &
    \end{tabular}
  \end{center}
  \caption{Ratio 
    $R^{V_{\mu}^{(i)}}(t;\boldvec{k}_{\pi},\boldvec{p}_B$
    for five combinations of $\boldvec{k}_{\pi}$ and
    $\boldvec{p}_B)$.
    Filled symbols represent the ratio for $V_4^{(0)}$, and
    open symbols are for $V_1^{(0)}$.
    Light quark is at $\kappa$=0.13711, and the heavy quark
    mass roughly corresponds to the $b$ quark mass,
    i.e. $aM_0$=3.0.
    }
  \label{fig:R}
\end{figure}

\begin{figure}
  \begin{center}
  \begin{tabular}{c}
    \psfig{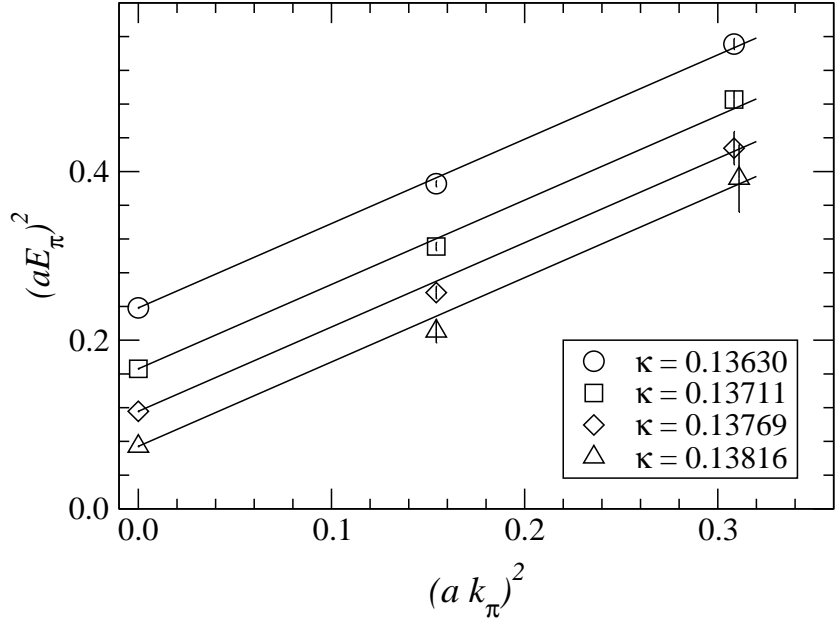}
  \end{tabular}
  \end{center}
  \caption{
    Dispersion relation for pion.
    The lines represent the continuum form
    (\ref{eq:dispersion_pi}). 
    }
  \label{fig:dispersion_pi}
\end{figure}

\begin{figure}
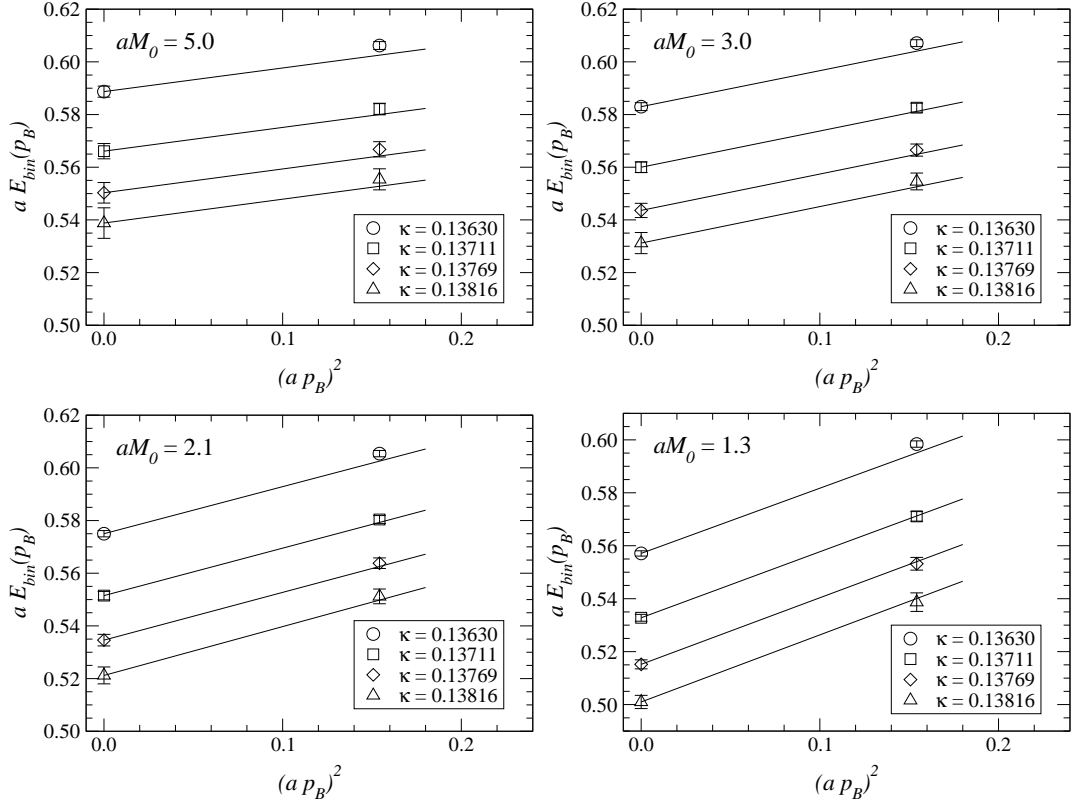

  \begin{center}
    \begin{tabular}{ll}
      \psfig{file=figure/dispersion_B_aM=5.0.eps,width=7cm,clip=,silent=} &
      \psfig{file=figure/dispersion_B_aM=3.0.eps,width=7cm,clip=,silent=} \\
      \psfig{file=figure/dispersion_B_aM=2.1.eps,width=7cm,clip=,silent=} &
      \psfig{file=figure/dispersion_B_aM=1.3.eps,width=7cm,clip=,silent=} 
    \end{tabular}
  \end{center}
  \caption{
    Dispersion relation for the heavy-light meson 
    at $aM_0$ = 5.0, 3.0, 2.1 and 1.3.
    The lines represent the non-relativistic form
    (\ref{eq:dispersion_B}) with perturbatively calculated
    meson mass $aM_B$.
    }
  \label{fig:dispersion_B}
\end{figure}

\begin{figure}
  \begin{center}
    \begin{tabular}{c}
      \psfig{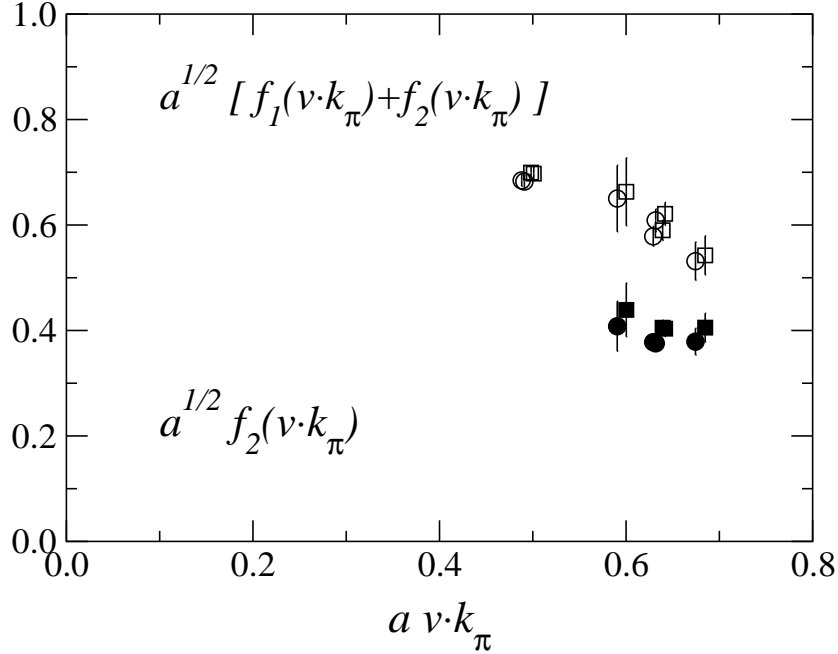}
    \end{tabular}
  \end{center}
  \caption{A typical plot of the form factors 
    $f_1(v\cdot k_{\pi})+f_2(v\cdot k_{\pi})$ (open symbols)
    and $f_2(v\cdot k_{\pi})$ (filled symbols) in the
    lattice unit.
    Parameters are $aM_0$=3.0, $\kappa$=0.13630, and
    $\alpha_V(1/a)$ (circles) or $\alpha_V(\pi/a)$ (squares)
    are used for the perturbative matching.
    The data for $\alpha_V(\pi/a)$ is slightly shifted in
    horizontal direction for clarity.
    }
  \label{fig:f1f2_typical}
\end{figure}

\begin{figure}
  \begin{center}
    \begin{tabular}{c}
      \psfig{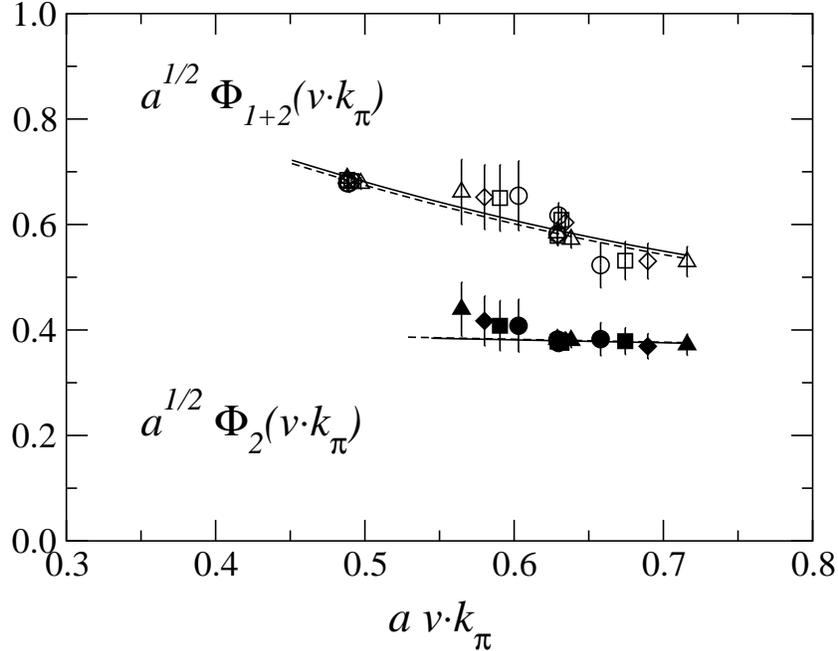}
    \end{tabular}
  \end{center}
  \caption{The renormalization group invariant form factors
    $\Phi_{1+2}(v\cdot k_{\pi})$ (open symbols)
    and $\Phi_2(v\cdot k_{\pi})$ (filled symbols) for different
    values of $aM_0$ with fixed light quark mass
    $\kappa$=0.13630. 
    Symbols denote the data at
    $aM_0$=5.0 (circles), 3.0 (squares), 2.1 (diamonds) and
    1.3 (triangles). 
    Solid and dashed lines show the fit 
    (\ref{eq:global_fit_f1+2})--(\ref{eq:global_fit_f2})
    for the heaviest ($aM_0$=5.0) and the lightest
    ($aM_0$=1.3) heavy quark masses respectively.
    }
  \label{fig:f1f2_heavy_dependence}
\end{figure}

\begin{figure}
  \begin{center}
    \begin{tabular}{c}
      \psfig{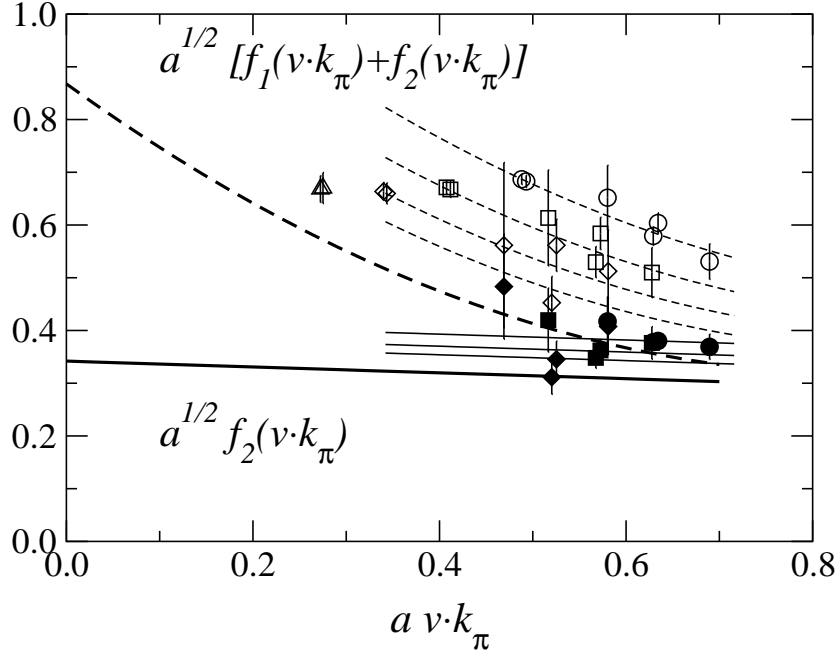}
    \end{tabular}
  \end{center}
  \caption{Light quark mass dependence of 
    $f_1(v\cdot k_{\pi})+f_2(v\cdot k_{\pi})$ (open symbols)
    and $f_2(v\cdot k_{\pi})$ (filled symbols) for
    $aM_0$=2.1.
    Symbols denote the data at $\kappa$=0.13630 (circles),
    0.13711 (squares), 0.13769 (diamonds) and 
    0.13816 (triangles). 
    Three thin solid lines, from above to below, show the
    fit (\ref{eq:global_fit_f2}) for $f_2(v\cdot k_{\pi})$
    with three $\kappa$ values, from heaviest to lightest.
    Four thin dashed lines, on the other hand, correspond to
    the fit (\ref{eq:global_fit_f1+2}) for the data of
    $f_1(v\cdot k_{\pi})+f_2(v\cdot k_{\pi})$
    for four values of $\kappa$.
    Thick lines represent the limit of physical light quark
    mass. 
    }
  \label{fig:f1f2_light_dependence}
\end{figure}

\begin{figure}
  \begin{center}
    \begin{tabular}{c}
      \psfig{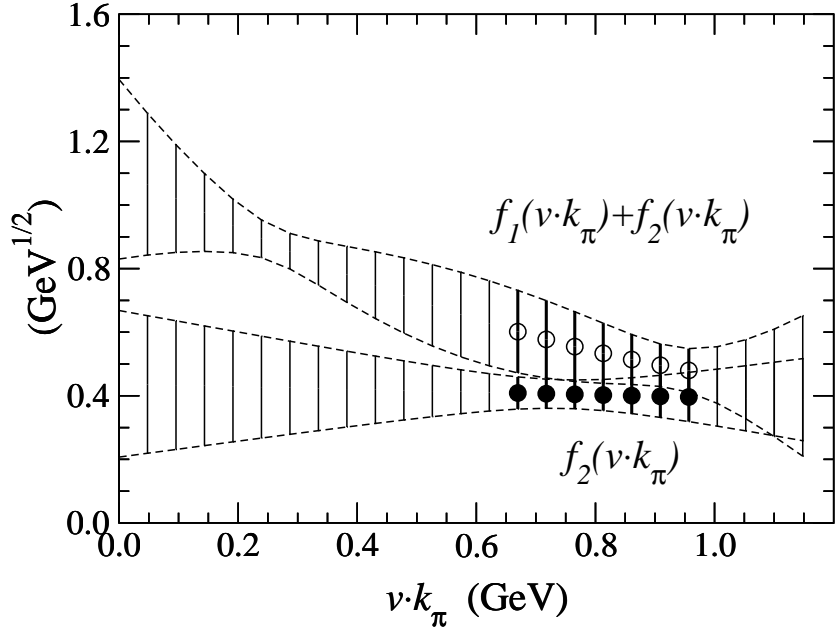}
    \end{tabular}
  \end{center}
  \caption{
    Form factors
    $f_1(v\cdot k_{\pi})+f_2(v\cdot k_{\pi})$ (open symbols)
    and $f_2(v\cdot k_{\pi})$ (filled symbols) at physical
    mass parameters.
    The points with symbols are obtained by interpolation in
    $v\cdot k_{\pi}$, while others involve extrapolations.
    }
  \label{fig:f1f2_physical}
\end{figure}

\begin{figure}
  \begin{center}
    \leavevmode\psfig{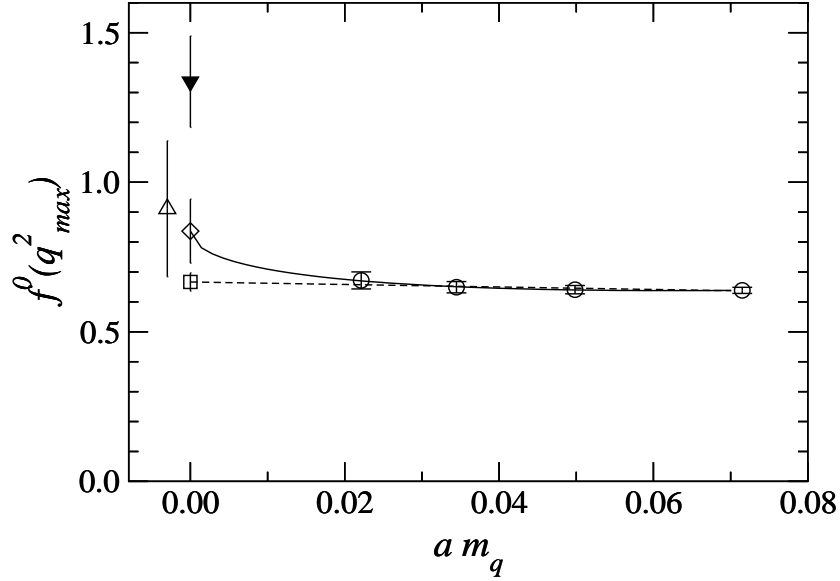}
  \end{center}
  \caption{
    Soft pion limit of 
    $f_0(q^2_{max})$ =
    $\frac{2}{\sqrt{m_B}}[f_1(v\cdot k_{\pi}) + f_2(v\cdot k_{\pi})]$
    at $aM_0$=3.0.
    The dashed line is a linear fit in $(am_{\pi})^2$, while
    the solid curve includes the term $(am_{\pi})$.
    A result of the fit (\ref{eq:global_fit_f1+2}) is given by an
    open triangle, which should be equal to $f_B/f_{\pi}$
    (filled triangle) in the soft pion theorem.
    }
  \label{fig:soft_pion_limit}
\end{figure}

\begin{figure}
  \begin{center}
    \begin{tabular}{c}
      \psfig{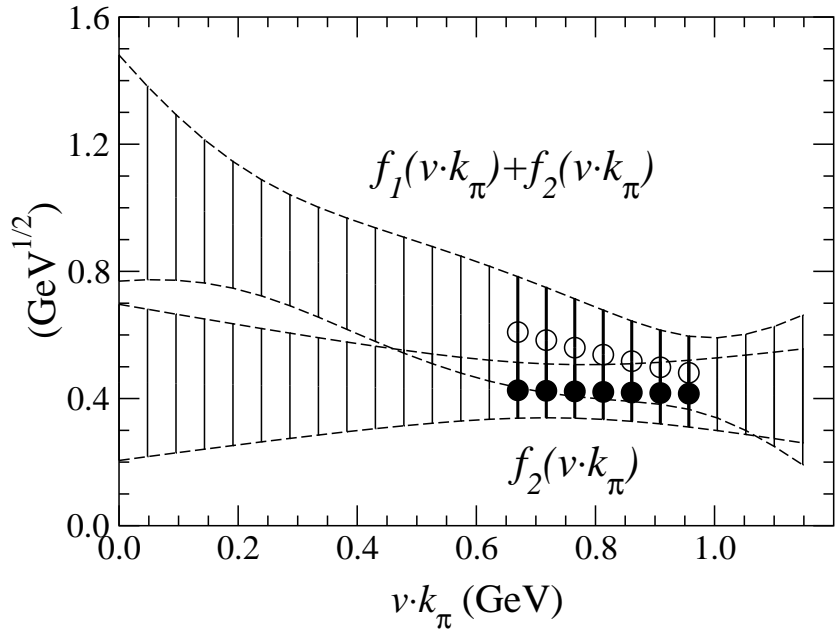}
    \end{tabular}
  \end{center}
  \caption{
    Same as Figure~\ref{fig:f1f2_physical}, but with
    estimated systematic errors.
    }
  \label{fig:f1f2_physical_with_syst}
\end{figure}

\begin{figure}
  \begin{center}
    \begin{tabular}{c}
      \psfig{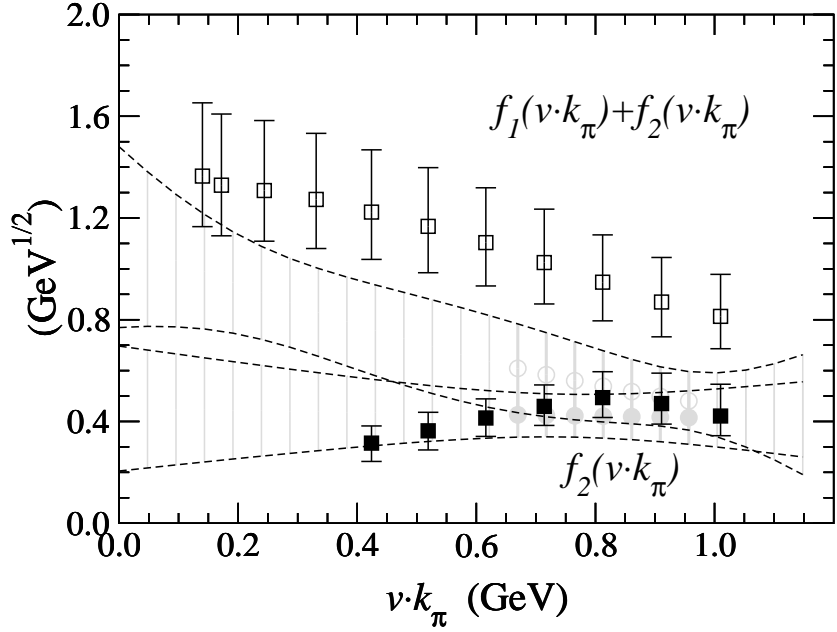}
    \end{tabular}
  \end{center}
  \caption{
    Form factors
    $f_1(v\cdot k_{\pi})+f_2(v\cdot k_{\pi})$ (open symbols)
    and $f_2(v\cdot k_{\pi})$ (filled symbols) at physical
    mass parameters.
    Squares represent the results of
    \protect\cite{Fermilab_01}, while our data presented in
    Figure~\protect\ref{fig:f1f2_physical} 
    is now plotted with gray symbols.
    }
  \label{fig:f1f2_fermilab}
\end{figure}

\begin{figure}
  \begin{center}
    \begin{tabular}{c}
      \psfig{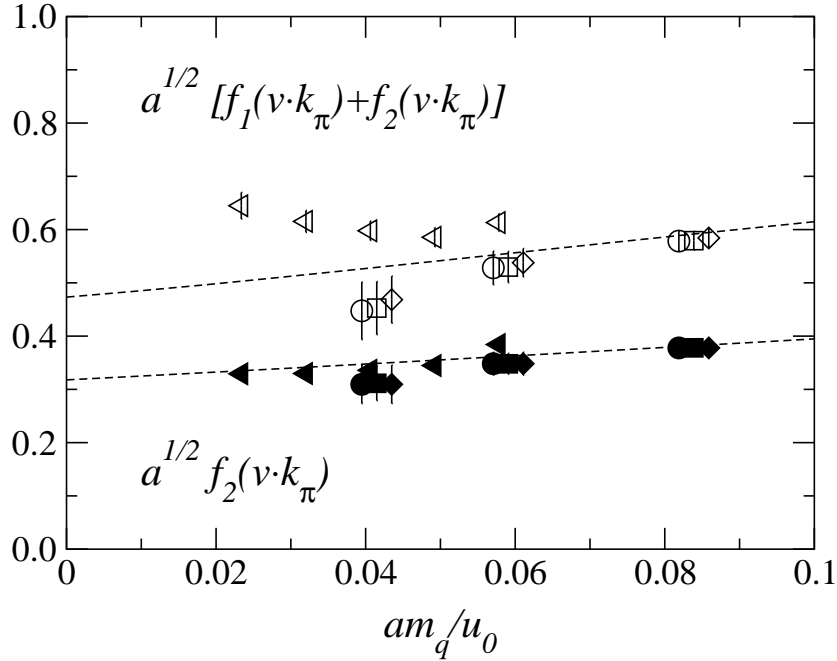}
    \end{tabular}
  \end{center}
  \caption{
    Form factors
    $f_1(v\cdot k_{\pi})+f_2(v\cdot k_{\pi})$ (open symbols)
    and $f_2(v\cdot k_{\pi})$ (filled symbols)
    for a fixed momentum configuration
    $a\boldvec{p}_B=(0,0,0)$ and $a\boldvec{k}_\pi=(1,0,0)$
    are plotted as a function of light quark mass $am_q/u_0$.
    Triangles are results of the Fermilab group
    \protect\cite{Fermilab_01} for a heavy quark mass close
    to the $b$ quark mass.
    Our results are shown for $aM_0$ = 3.0 (circles), 2.1
    (squares) and 1.3 (diamonds).
    Squares and diamonds are shifted in the horizontal
    direction for clarity.
    Lines show the global fit (\ref{eq:global_fit_f1+2}) and
    (\ref{eq:global_fit_f2}).
    }
  \label{fig:mq_dependence_fermilab}
\end{figure}

\begin{figure}
  \begin{center}
    \begin{tabular}{c}
      \psfig{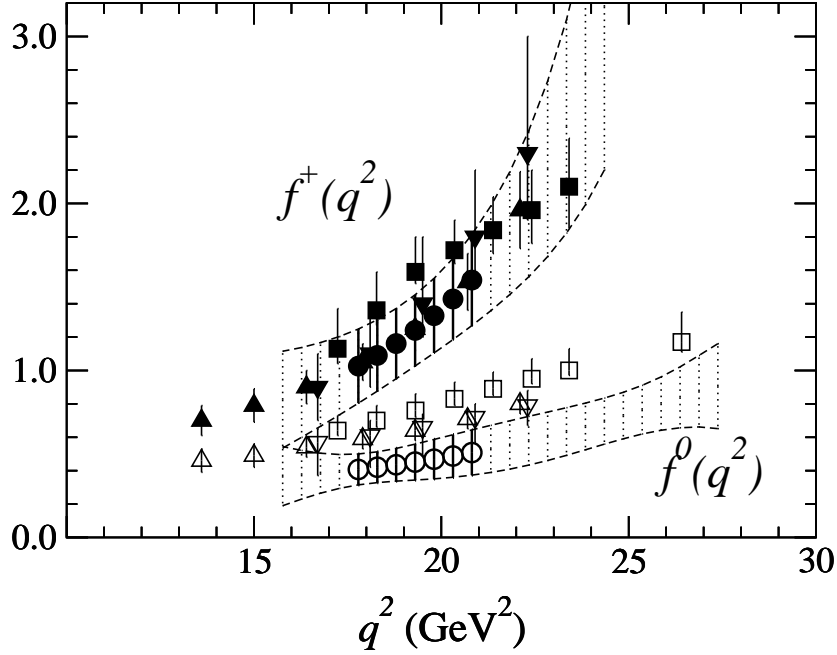}
    \end{tabular}
  \end{center}
  \caption{
    Comparison of the results for the form factors
    $f^+(q^2)$ (filled symbols) and $f^0(q^2)$ (open
    symbols). 
    Data are from APE \protect\cite{APE_00} (up trianlgles),
    UKQCD \protect\cite{UKQCD_00} (down triangles) and
    Fermilab \protect\cite{Fermilab_01} (squares).
    Our results are plotted by circles and error bands
    are shown by dashed lines.
    }
    \label{fig:fpf0_comparison}
\end{figure}

\begin{figure}
  \begin{center}
    \begin{tabular}{c}
      \psfig{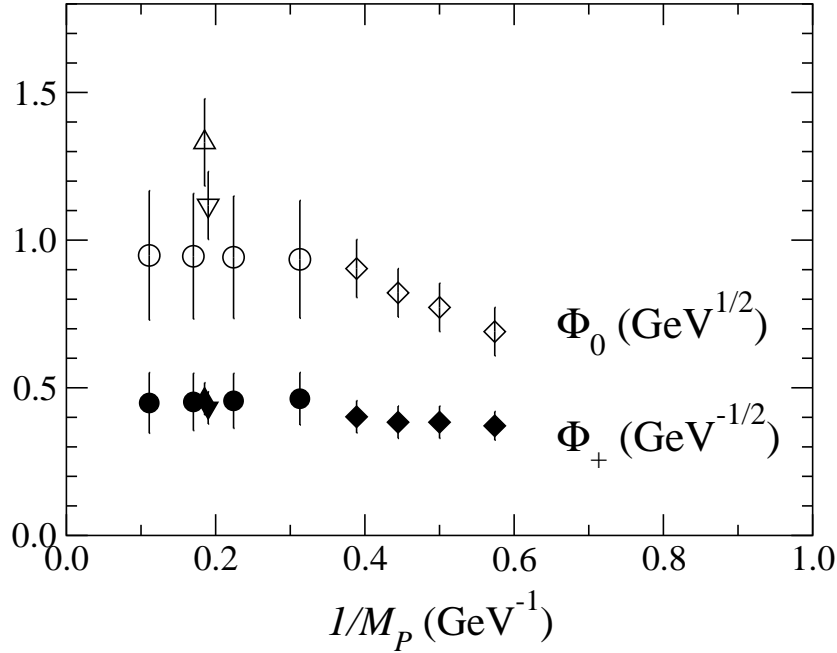}
    \end{tabular}
  \end{center}
  \caption{
    $1/M_P$ dependence of the form factors
    $\Phi_+\equiv (\alpha_s(M_P)/\alpha_s(M_B)^{-2/11}f^+/\sqrt{M_P}$ 
    (filled symbols)
    $\Phi_0\equiv (\alpha_s(M_P)/\alpha_s(M_B)^{-2/11}f^0\sqrt{M_P}$ 
    $f^+(q^2)$ (open symbols)
    at a fixed $v\cdot k_{\pi}$ ($=$ 0.845 GeV).
    Simulation results from the APE collaboration
    \protect\cite{APE_00} are shown by diamonds, and their
    linear and quadratic extrapolation to the $B$ meson mass
    is plotted by down and up triangles, respectively.
    Our results are given by circles.
    }
    \label{fig:phi_comparison}
\end{figure}

\begin{figure}
  \begin{center}
    \leavevmode\psfig{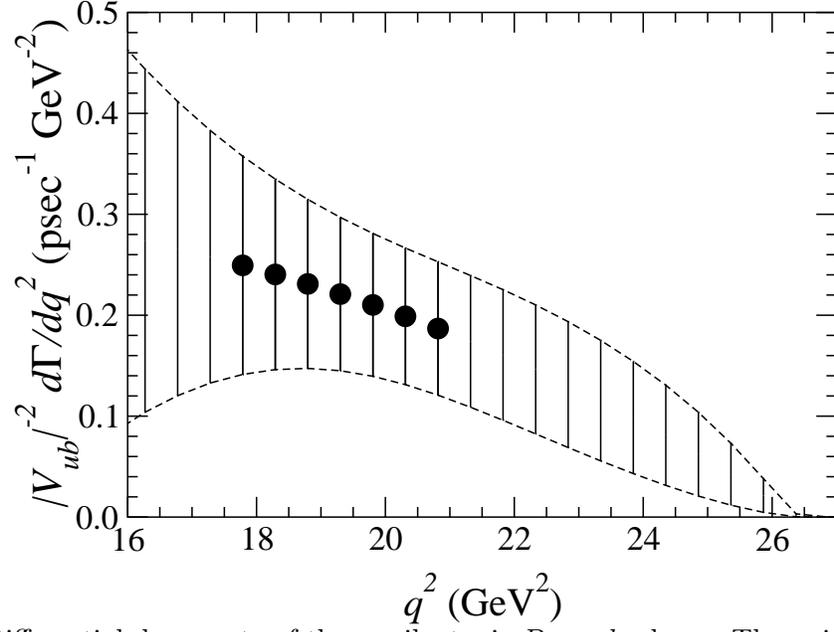}
    \caption{
      Differential decay rate of the semileptonic
      $B\rightarrow\pi l\nu$ decay.
      The points with symbols are obtained by interpolation
      in $v\cdot k_{\pi}$, while others involve
      extrapolations. 
      }
    \label{fig:diff_decayrate}
  \end{center}
\end{figure}
